\documentclass{aastex631}

\usepackage{array}
\usepackage{threeparttable}
\usepackage{multirow}
\usepackage{booktabs}
\usepackage{amsmath}
\usepackage{graphicx}
\usepackage{anyfontsize}
\usepackage{rotating} 
\usepackage{longtable} 
\usepackage{rotfloat} 

\begin{document}

\title{Physical parameters of 12201 ASAS-SN contact binaries determined by the Neural Network}

\author[0000-0003-3590-335X]{Kai Li}
\affiliation{Shandong Key Laboratory of Optical Astronomy and Solar-Terrestrial Environment, School of Space Science and Technology, Institute of Space Sciences, Shandong University, Weihai, Shandong, 264209, China}

\author[0009-0005-0485-418X]{Li-Heng Wang}
\affiliation{Shandong Key Laboratory of Optical Astronomy and Solar-Terrestrial Environment, School of Space Science and Technology, Institute of Space Sciences, Shandong University, Weihai, Shandong, 264209, China}
\correspondingauthor{Kai Li}
\email{kaili@sdu.edu.cn}

\begin{abstract}
In the era of astronomical big data, more than one million contact binaries have been discovered. Traditional approaches of light curve analysis are inadequate for investigating such an extensive number of systems. This paper builds on prior research to present an advanced Neural Network model combined with the Markov Chain Monte Carlo algorithm and including spot parameters. This model was applied to 12785 contact binaries selected from All-Sky Automated Survey for Supernovae. By removing those with goodness of fit less than 0.8, we obtained the physical parameters of 12201 contact binaries. Among these binaries, 4332 are A-subtype systems, while 7869 are W-type systems, and 1594 systems have mass ratios larger than 0.72 (H-subtype system). A statistical study of the physical parameters was carried out, and we found that there are two peaks in the mass ratio distribution and that the probability of the presence of spot is about 50\%. In addition, the differences in flux between the two light maxima are from $-$0.1 to 0.1. As the orbital period and the temperature of the primary component decrease, the difference between the two light maxima becomes more pronounced. Based on the relationships between transfer parameter and luminosity ratio, as well as between luminosity ratio and mass ratio, we found that A-, W-, and H-type contact binaries are distributed in distinct regions. 

\end{abstract}


\section{Introduction} \label{sec:intro}

The contact binary is a close binary whose components are usually late type stars sharing a common envelope \citep{1968ApJ...151.1123L,1968ApJ...153..877L}. They are the most numerous close binary stars, and roughly one of every five hundred main sequence stars is a contact binary in our Galaxy \citep{2007MNRAS.382..393R}. They are very important tools for determining precise physical parameters of stars \citep{2001AJ....122.1974R} and for determining the distance (e.g., \citealt{1994PASP..106..462R,2016ApJ...832..138C,2018ApJ...859..140C}). Although contact binaries have been analyzed for more than eighty years \citep{1941ApJ....93..133K}, there are still many open questions, such as the orbital period cut-off \citep{1992AJ....103..960R,2019MNRAS.485.4588L,2020AJ....159..189L}, the instability mass ratio \citep{1995ApJ...444L..41R,2024SerAJ.208....1A,2022AJ....164..202L,2024NatSR..1413011Z,2024A&A...692L...4L}, the O'Connell effect \citep{1951PRCO....2...85O}, and their formation and evolution \citep{1988ASIC..241..345G,1994ASPC...56..228B,2017RAA....17...87Q}. In order to solve these issues, physical parameters of an increasing number of contact binaries are needed.

Thanks to ground- and space- based photometric surveys, more than one million contact binaries have been detected (e.g., \citealt{2018MNRAS.477.3145J,2020ApJS..249...18C,2022ApJS..258...16P,2023A&A...674A..16M}). The light curves (LCs) of these surveys are released and can be used to determine the physical parameters of contact binaries. The traditional method using the Wilson-Devinney (W-D) code \citep{1971ApJ...166..605W, 1979ApJ...234.1054W, 1990ApJ...356..613W} or the PHysics Of Eclipsing BinariEs (Phoebe; \citealt{2005ApJ...628..426P,2016ApJS..227...29P}) is time-consuming and impractical for calculating the parameters of such a large amount of contact binaries. Recently, \cite{2022AJ....164..200D} developed a model that can rapidly derive the physical parameters of a contact binary based on machine learning methods and the Markov Chain Monte Carlo (MCMC) algorithm \citep{2019JOSS....4.1864F} using synthetic LCs generated by Phoebe, and they have applied this method to model LCs from the Transiting Exoplanet Survey Satellite (TESS; \citealt{2023MNRAS.525.4596D}), the All-Sky Automated Survey for SuperNovae (ASAS-SN; \citealt{2024ApJS..271...32L} and the Catalina Sky Survey (CSS; \citealt{2024ApJS..273...31W}). Although they derived physical parameters for a large number of contact binaries, they did not include star spots. Most contact binaries are late type stars and exhibit asymmetric LCs (this is called the O'Connell effect; \citealt{1951PRCO....2...85O}), and star spots on one of the components are needed to model the LCs, and adding a spot will affect the physical parameters of a contact binary \citep{2007AJ....134.1475Q,2021PASP..133h4202L}. This paper constructs a model that can include spot parameters based on a Neural Network (NN) and Phoebe generated LCs and applies it to the EW-type eclipsing binaries identified by ASAS-SN. The constructed NN models and the physical parameters of the contact binaries are available at China-VO\dataset[(DOI: 10.12149/101561)]{https://doi.org/10.12149/101561}. Our model significantly improves computational efficiency and is particularly useful for processing the large amount of data from photometric surveys.  By including star spots in our model, we can derive more precise physical parameters for contact binaries exhibiting the O'Connell effect, advancing our understanding of the structure, evolution, and energy transfer mechanisms in contact binaries.

\section{Data selection} \label{sec:data}
The ASAS-SN Catalog of Variable Stars \citep{2014ApJ...788...48S,2019MNRAS.485..961J} has identified 78503 EW-type eclipsing binaries. We used two criteria to select samples to analyze. First, the class probability should be no less than 0.99; Second, the Lafler–Kinman string length (LKSL; \citealt{1965ApJS...11..216L,2002A&A...386..763C}) statistic should be smaller than 0.1. We downloaded the flux LCs of the resulting 12785 targets, and removed the data points below the 1st percentile and above the 99th percentile. To determine the primary eclipse time of the binaries, we performed the following procedures. Firstly, the data were shifted into one period using the orbital period determined by \cite{2019MNRAS.485..961J} and an initial reference time. Secondly, a sine curve was fit to the data to determine the time of the LC maximum. Thirdly, using this time as a reference time, we selected the phased data below the median of the flux and fit a parabola to find the minimum of the LC. Centering on the minimum value, we sequentially selected data points below the median flux on both sides and used the \cite{1956BAN....12..327K} method to calculate the time of the minimum. Using these primary eclipse times, we calculated the phases for all targets. Finally, we normalized the flux by setting the value at phase 0.25 to unity.

\section{The Neural Network models} \label{sec:model}
The NN model and the MCMC algorithm were used to determine the physical parameters of the contact binaries. Synthetic LCs generated by Phoebe were used for training. The following parameters are considered in the training, the temperature of the primary component ($T_1$, [4000K, 8000K]), the temperature ratio ($T_2/T_1$, [0.7,1.2]), the mass ratio ($q$, [0,1]), the orbital inclination ($i$, [$30^\circ,90^\circ$]), the fill-out factor ($f$, [0,1]), and third light ($l_3$, [0,0.6]). If the LCs are asymmetric, spot parameters should also be added. There are four parameters for a spot, latitude ($\theta$), longitude ($\lambda$), angular radius ($r_s$), and relative temperature ($T_s$). If we set all of these as adjustable parameters, it would lead to an exponential increase in computations. Therefore, we fixed $\theta$ as $90^\circ$, $\lambda$ as $90^\circ$ or $270^\circ$ (this is due to the brightnesses of the two light maxima, if the primary maximum is brighter than the secondary maximum, $\lambda$ is set as $90^\circ$, if the primary maximum is fainter than the secondary maximum, $\lambda$ is set as $270^\circ$), $r_s$ as $20^\circ$, $T_s$ was set as an adjustable parameter, ranging from 0.6 to 1. The \cite{2004A&A...419..725C} atmospheric model and the Johnson V  filter were used. The gravity darkening and bolometric albedo coefficients were set as $g=0.32$ \citep{1967ZA.....65...89L} and $A=0.5$ \citep{1969AcA....19..245R} when the temperature is less than 7200 K, while the two parameters were set to unity when the temperature is higher than 7200 K.

\begin{figure*}
\centering
\includegraphics[width=0.32\textwidth]{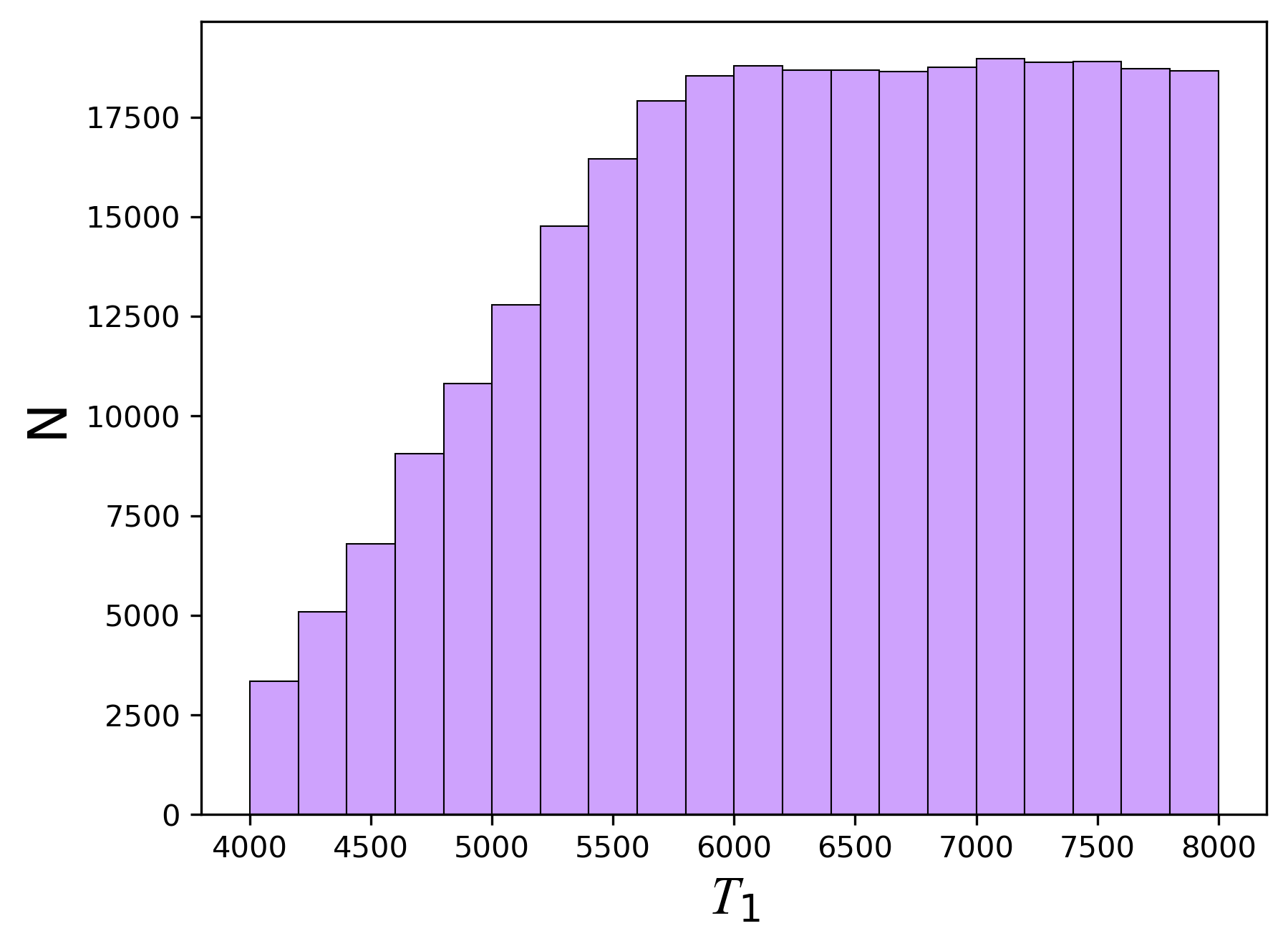}
\includegraphics[width=0.32\textwidth]{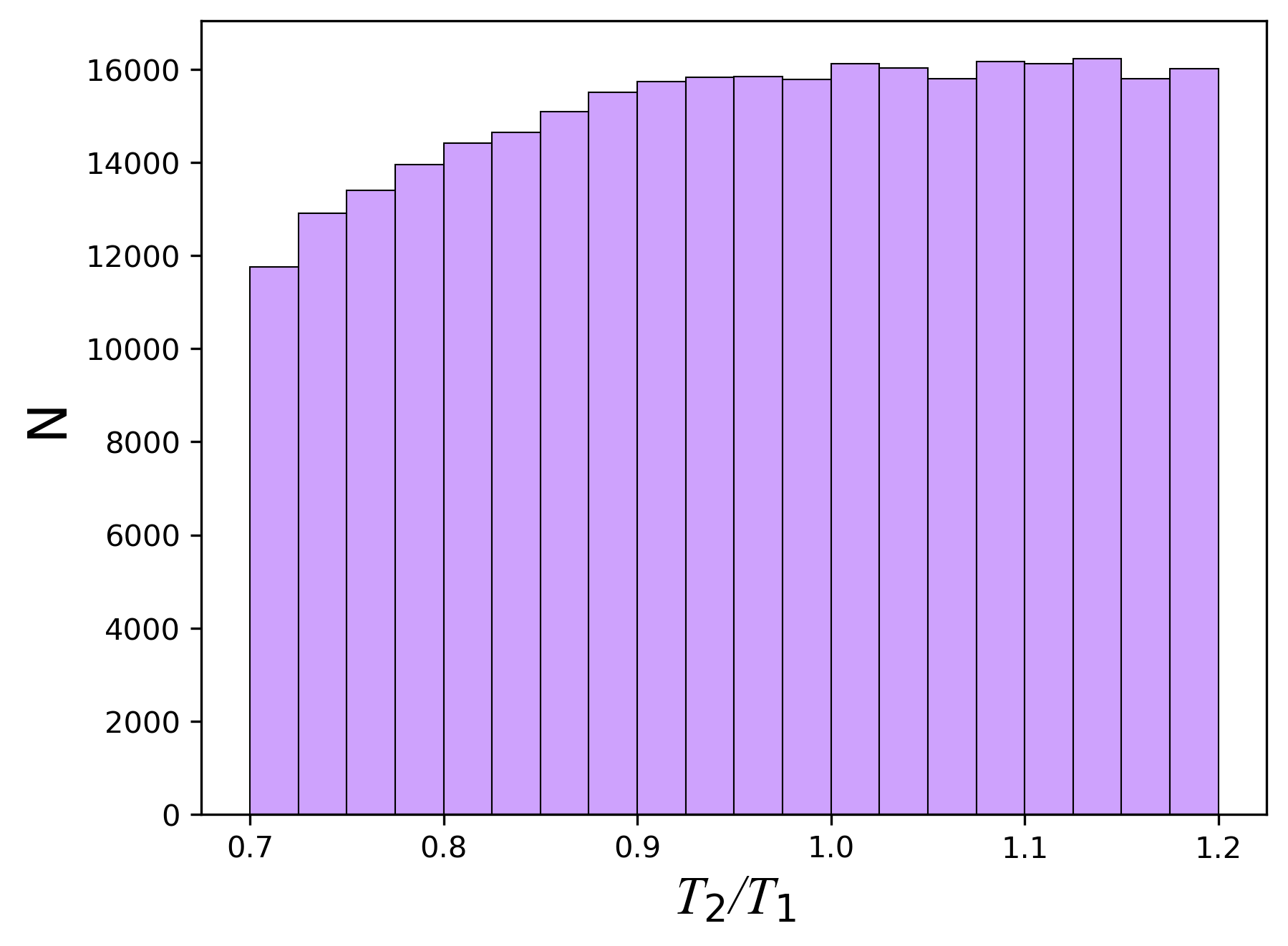}
\includegraphics[width=0.32\textwidth]{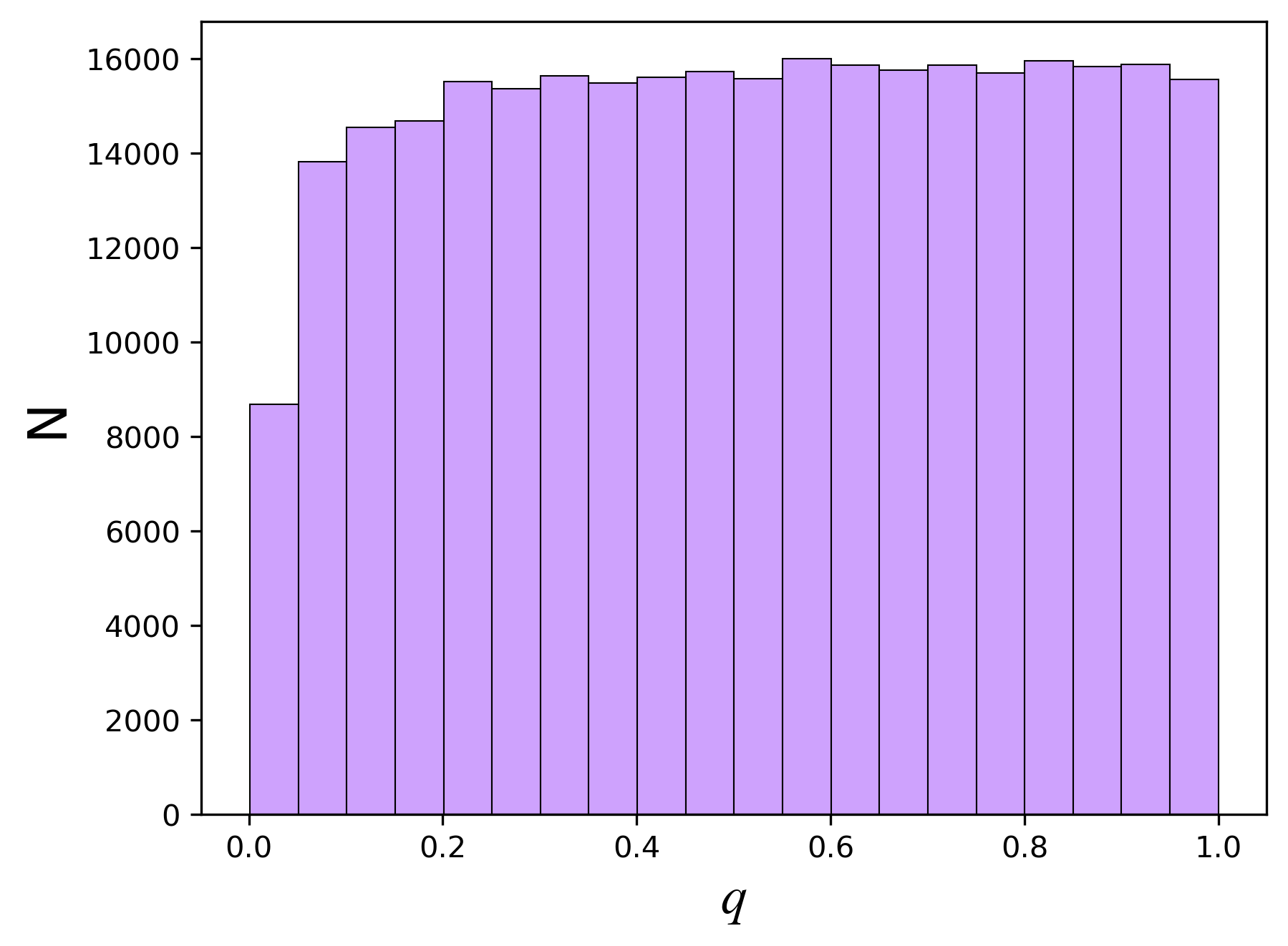}
\includegraphics[width=0.32\textwidth]{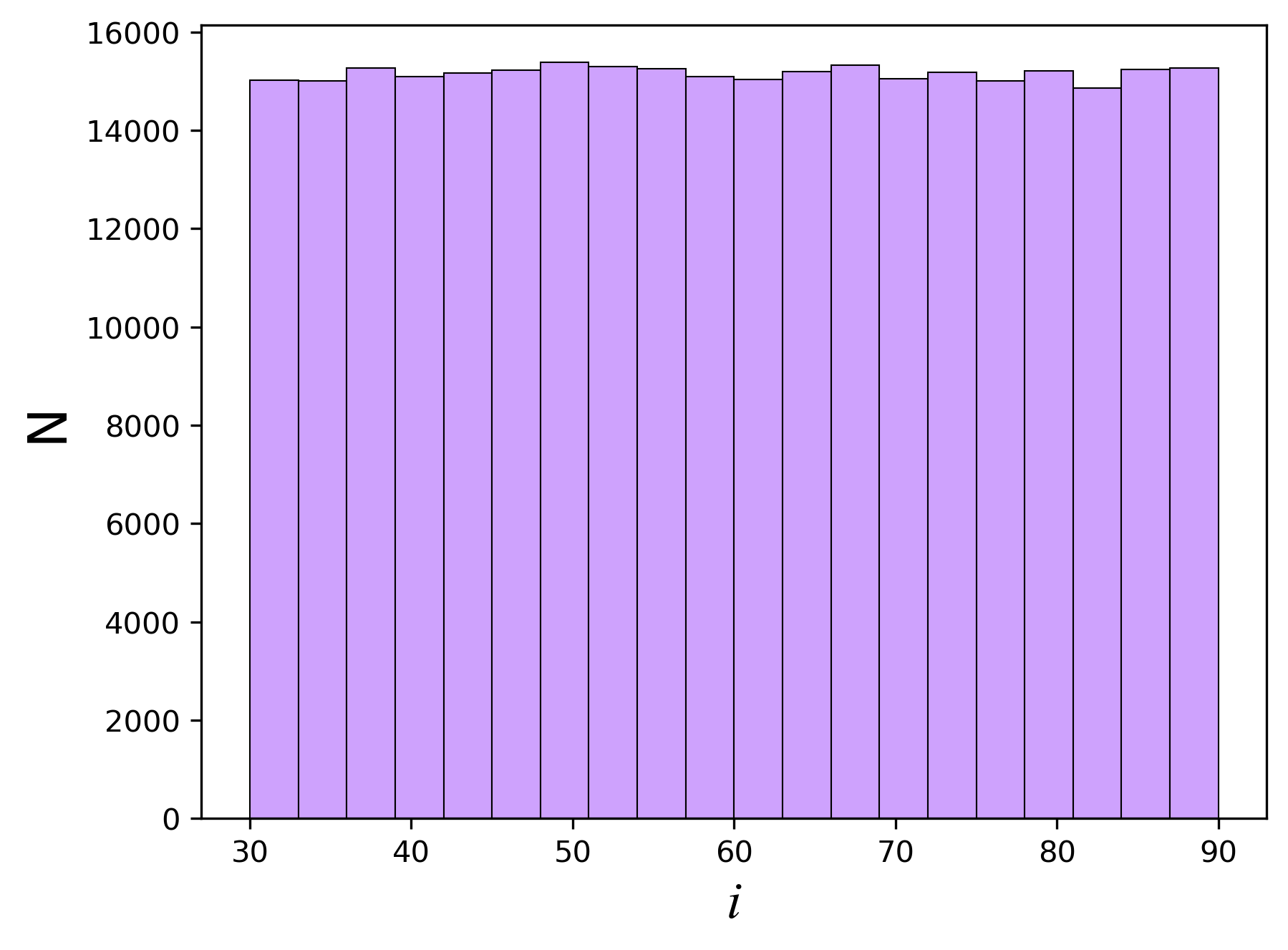}
\includegraphics[width=0.32\textwidth]{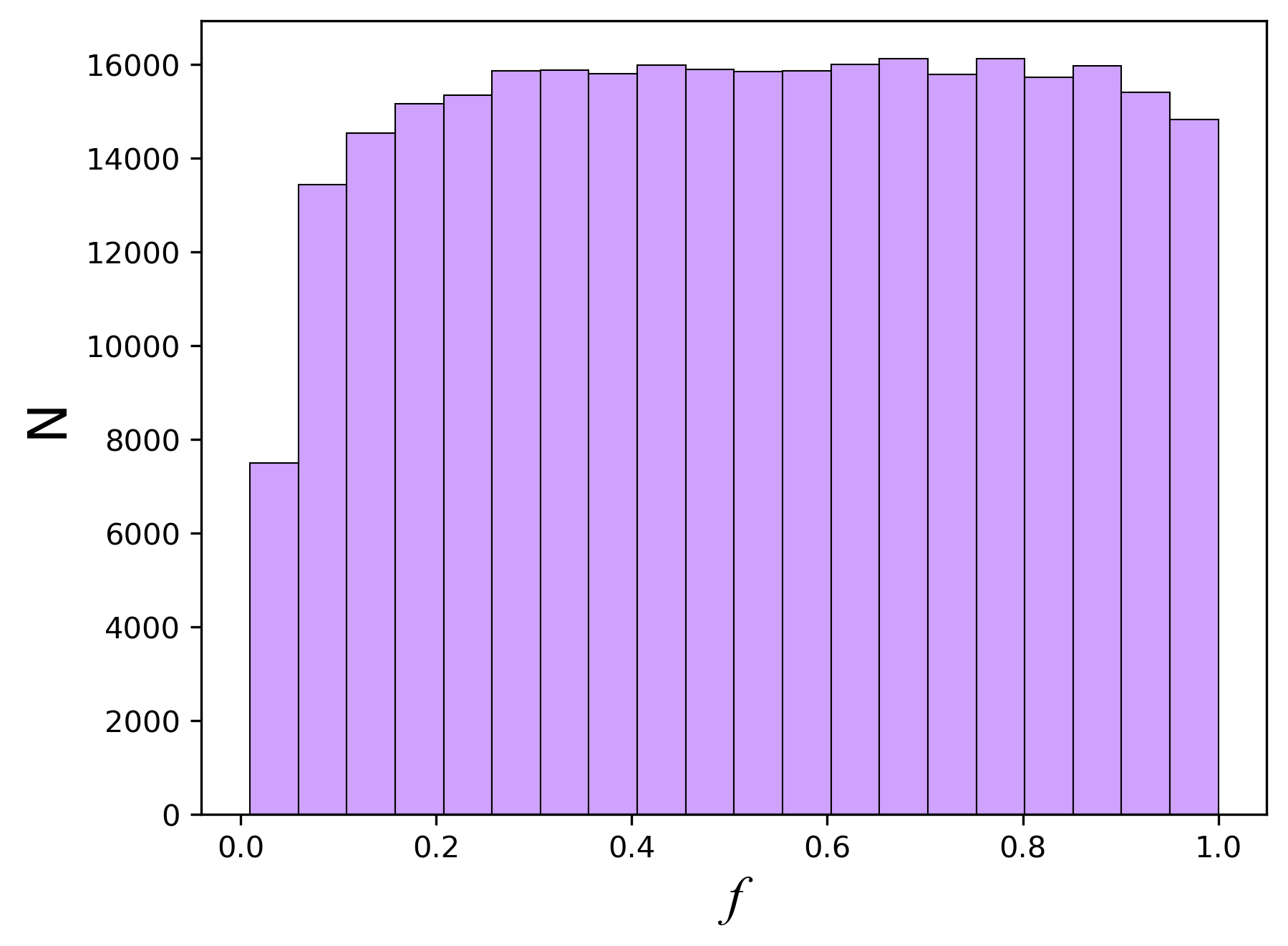}
\includegraphics[width=0.32\textwidth]{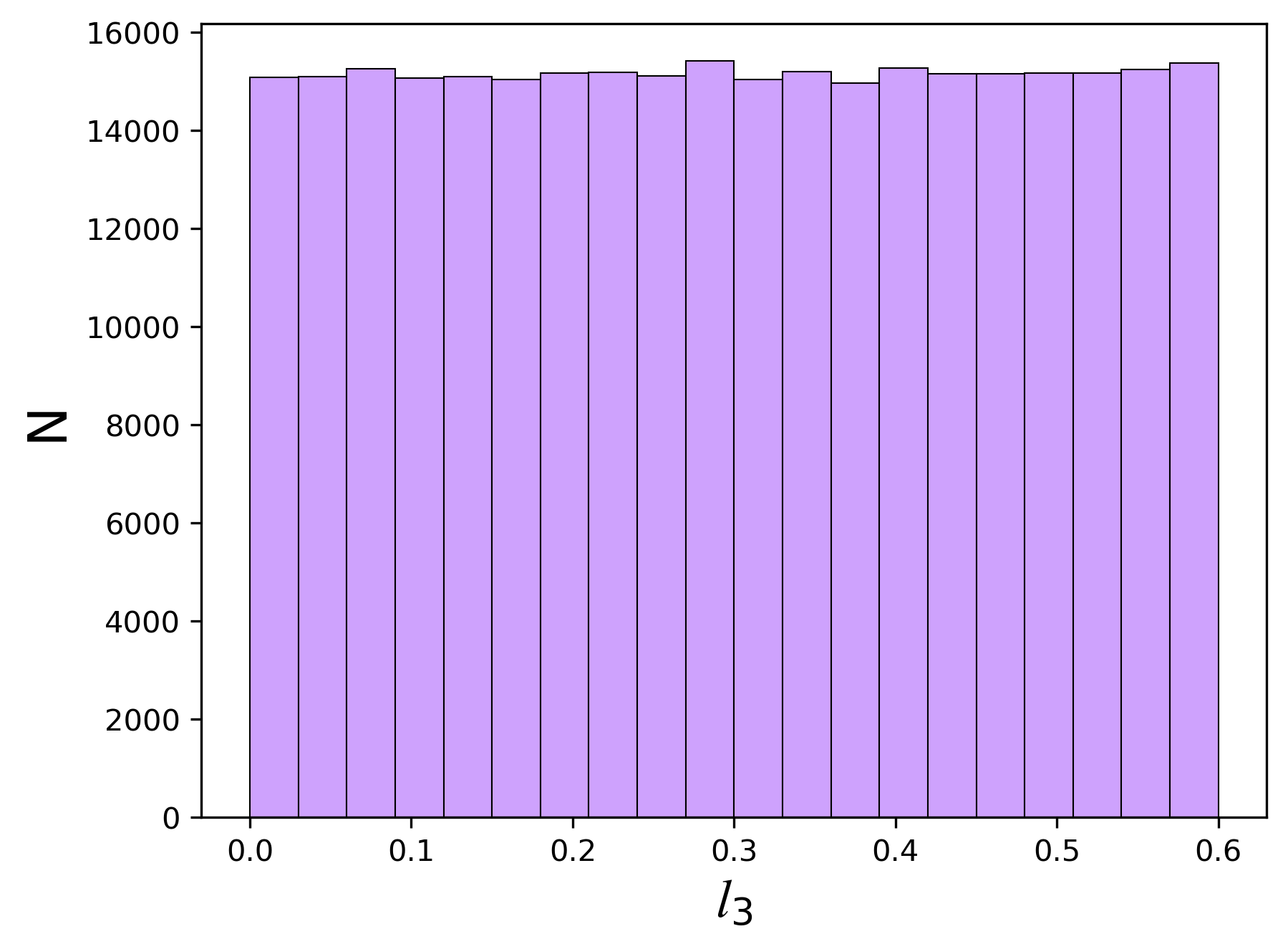}
\includegraphics[width=0.32\textwidth]{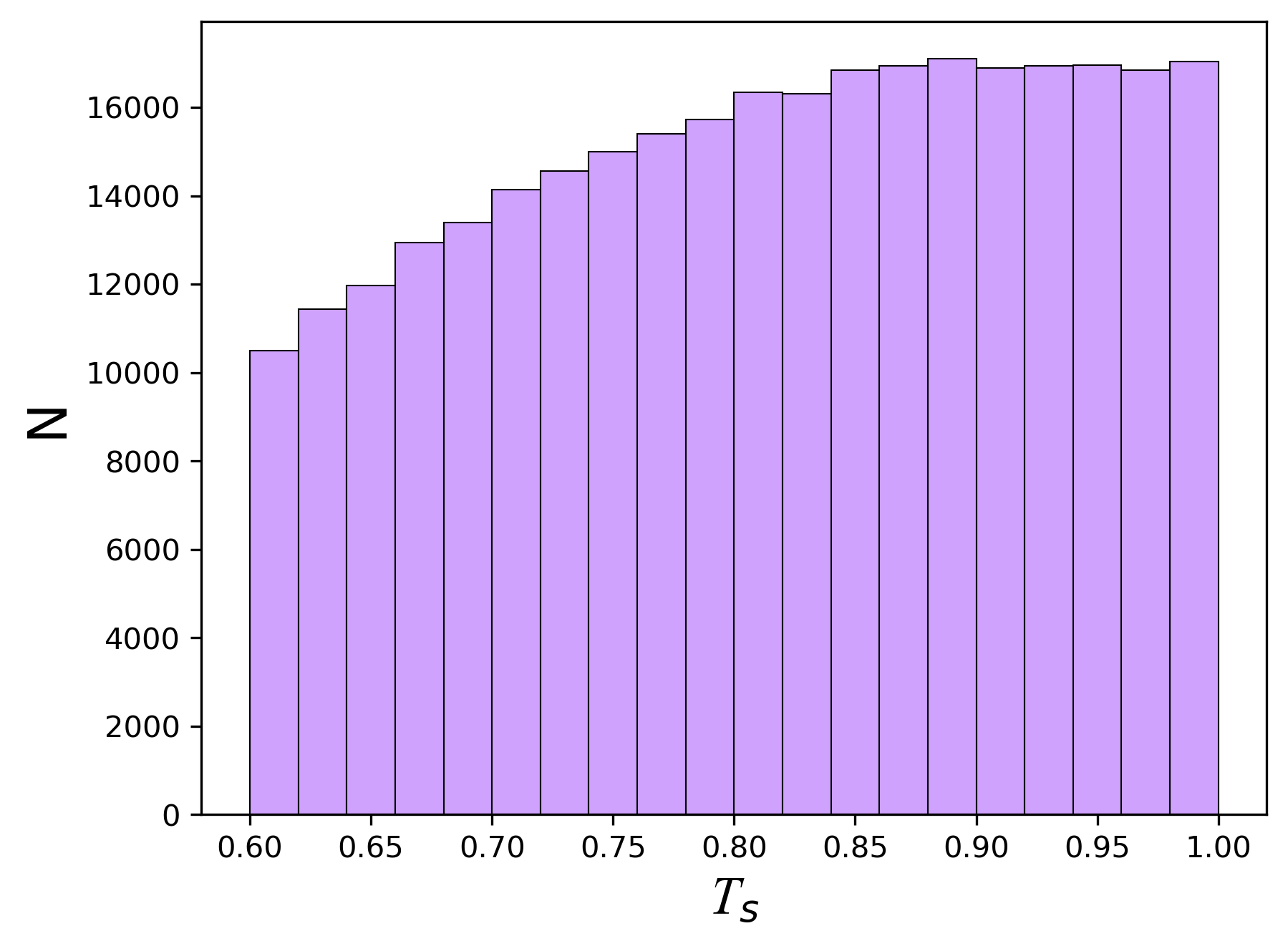}
\caption{The parameter distributions of Model 5.}
\label{Fig1}
\end{figure*}

\begin{figure*}
\centering
\includegraphics[width=0.5\textwidth]{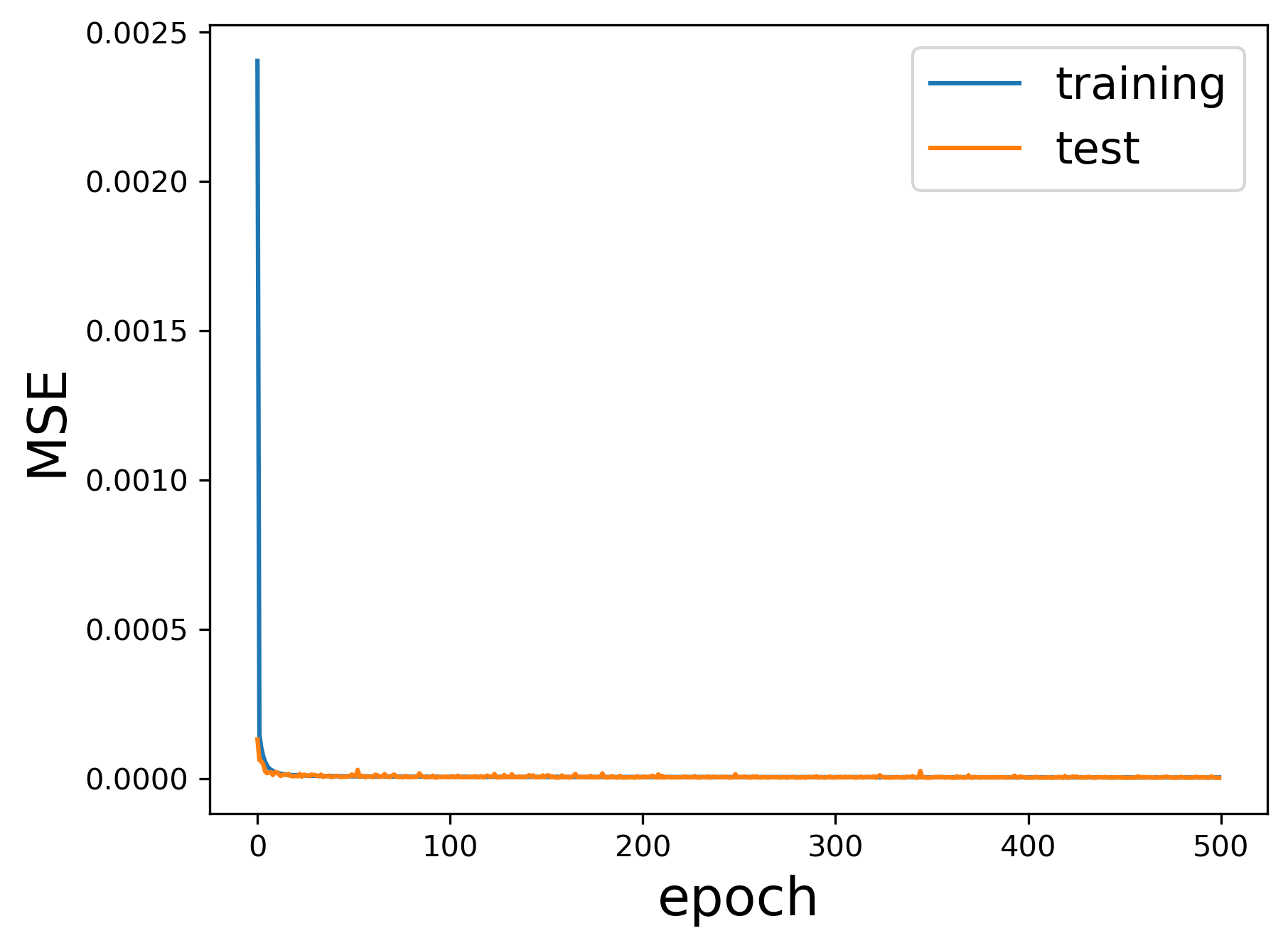}

\caption{Loss function of the training and test sets of Model 5.}
\label{Figt}
\end{figure*}

We constructed six NN models with different architectures: a model without third light or spot parameters (Model 1), a model with spot parameters but without third light (Model 2 for $\lambda=90^\circ$ and Model 3 for $\lambda=270^\circ$), a model with third light but without spot parameters (Model 4), and a model with third light and spot parameters (Model 5 for $\lambda=90^\circ$ and Model 6 for $\lambda=270^\circ$). We uniformly sample over the parameters ($T_1$, $T_2/T_1$, $q$, $i$, $f$, $l_3$, $T_s$) to generate LCs with 100 phased points with Phoebe. There are 200000 LCs for Model 1, and 400000 for the other models. During LC generation, some parameters that do not conform to the physical model of contact binaries were excluded. The parameter distributions of Model 5 are shown in Figure \ref{Fig1} for example. Using similar parameter scales can lead to better training results, so we rescale some of the parameters to $T_1$/6000,  $i$/90, and $l_3$/0.6. In order to obtain the best numbers of hidden layers and neurons, we conducted training with numbers of hidden layers set to 2, 3, 4, or 5, and with numbers of neurons set to 100, 150, 200, or 500 for each model. We divided the number of dataset into an 80\% training set and a 20\% test set. $batch\_size=100$ and $epoch=500$ were used for training, the mean square error (MSE) was used as the loss function, and ReLU \citep{2015arXiv150201852H} was used as the activation function. The Adam optimizer \citep{2014arXiv1412.6980K} was used for training the model. Figure \ref{Figt} shows the loss function of the training and test sets of Model 5. After completion of the training, we randomly generated 1,000 light curves using both PHOEBE and our model. We then calculated the average goodness of fit ($R^2$). The six combinations of hidden layers and neurons number with the highest average $R^2$ were selected as the final models and are listed in Table \ref{tab:NN}.

\begin{table}
\centering
\small
\caption{ The six NN models} \label{tab:NN}
\begin{tabular}{cccc}
\hline
Model &Hidden layers& Neurons& R$^{2*}$  \\
\hline
1 & 5& 300 & 0.9996\\
2 & 4& 200 & 0.9994\\
3 & 5& 100 & 0.9989\\
4 & 5& 100 & 0.9989 \\
5 & 4& 400 & 0.9992 \\
6 & 5& 400 & 0.9988 \\
\hline
$^*$ R$^2$ means the goodness of fit.
\end{tabular}
\end{table}

\section{Physical parameter determination using MCMC} \label{sec:physical parameter}
Based on our six NN models, we used the MCMC algorithm to determine the physical parameters of the contact binaries. The brightness difference ($\Delta m=Max. I-Max. II$) between the two light maxima is calculated for all the targets, if $\left| \Delta m\right|$ is less than 0.01, the LC will be treated as symmetric; otherwise, the LC is treated as asymmetric. Symmetric LCs were analyzed using Models 1 and 4. For asymmetric LCs, if $\Delta m$ is larger than $+0.01$, Models 2 and 5 were used, while $\Delta m$ is smaller than $-0.01$, Models 3 and 6 were used. The temperature of the primary component was taken from the Gaia DR3 catalog \citep{2016A&A...595A...1G,2023A&A...674A...1G}.  If there is no temperature in Gaia DR3, we used the period–temperature relationship obtained by \cite{2020MNRAS.493.4045J} to estimate it.

\begin{figure*}
\centering
\includegraphics[width=0.45\textwidth]{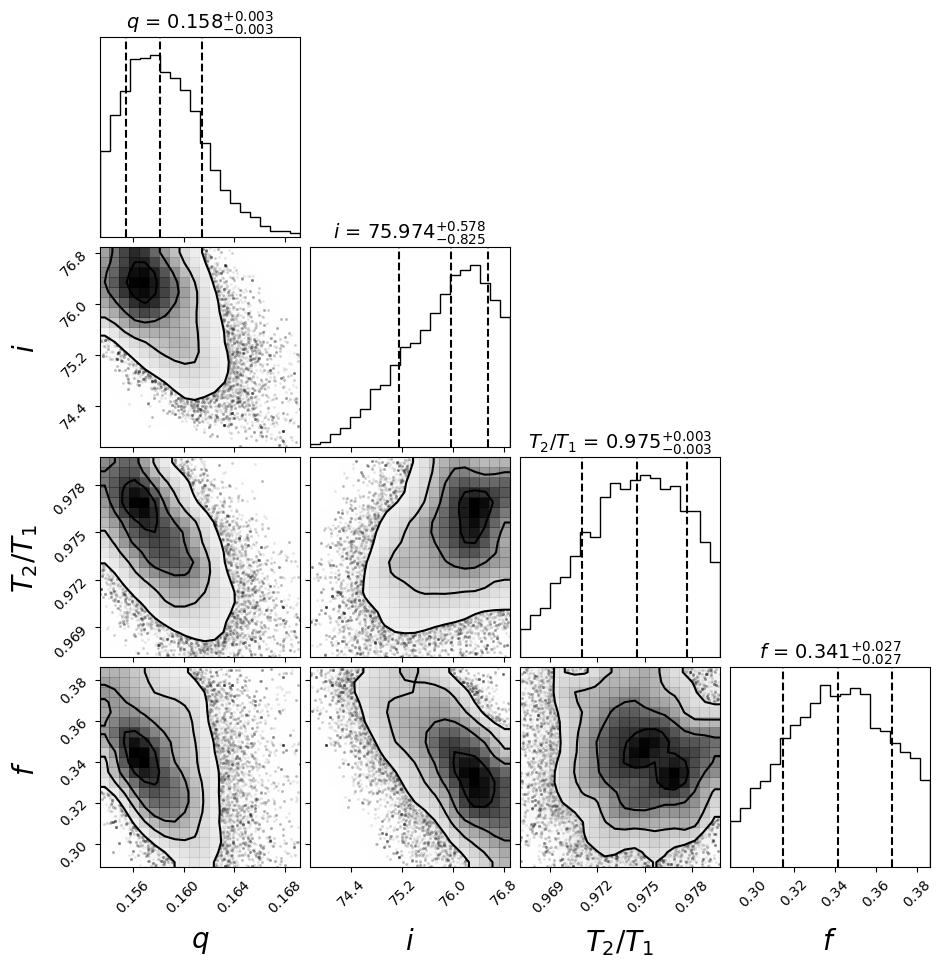}
\includegraphics[width=0.45\textwidth]{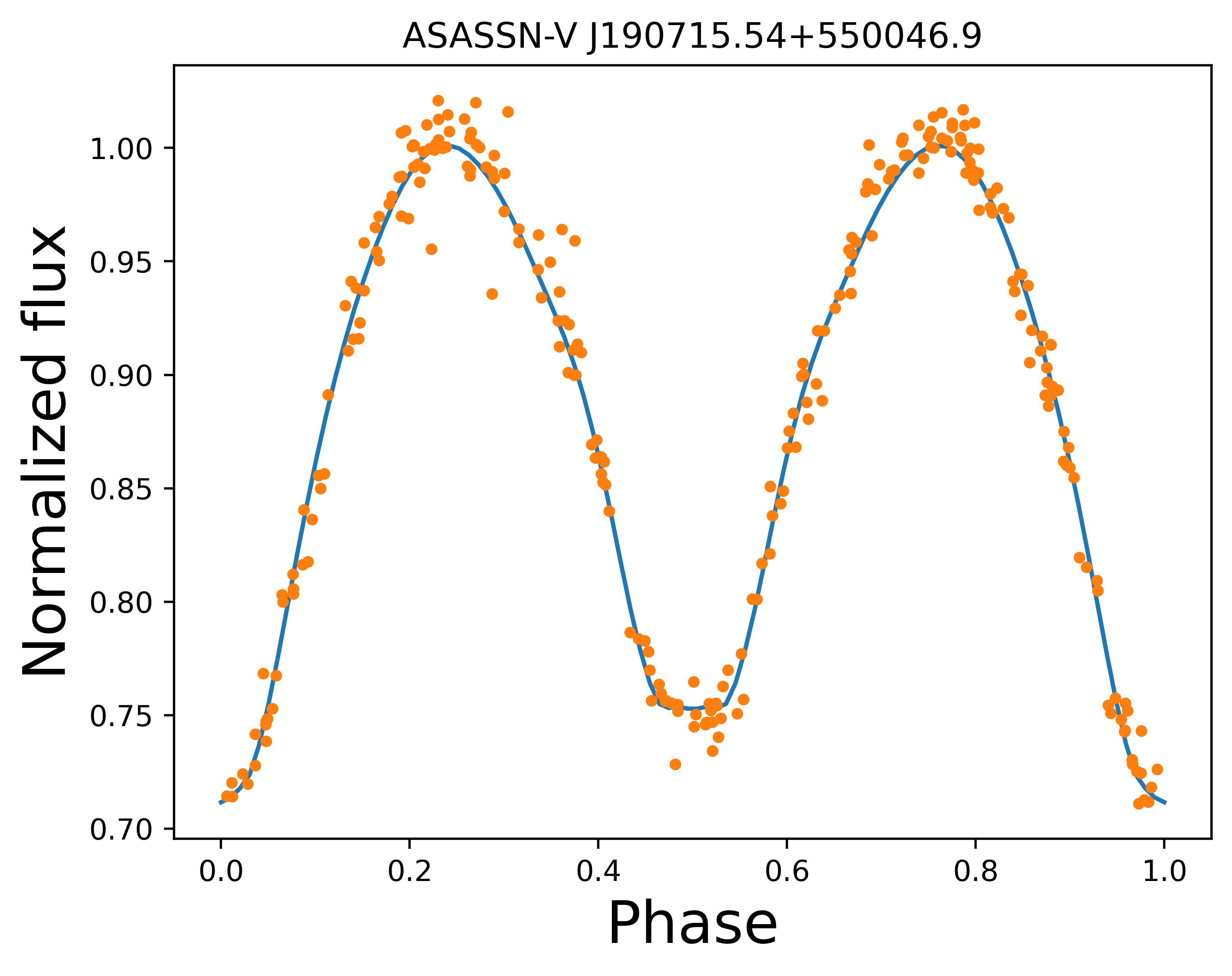}
\includegraphics[width=0.45\textwidth]{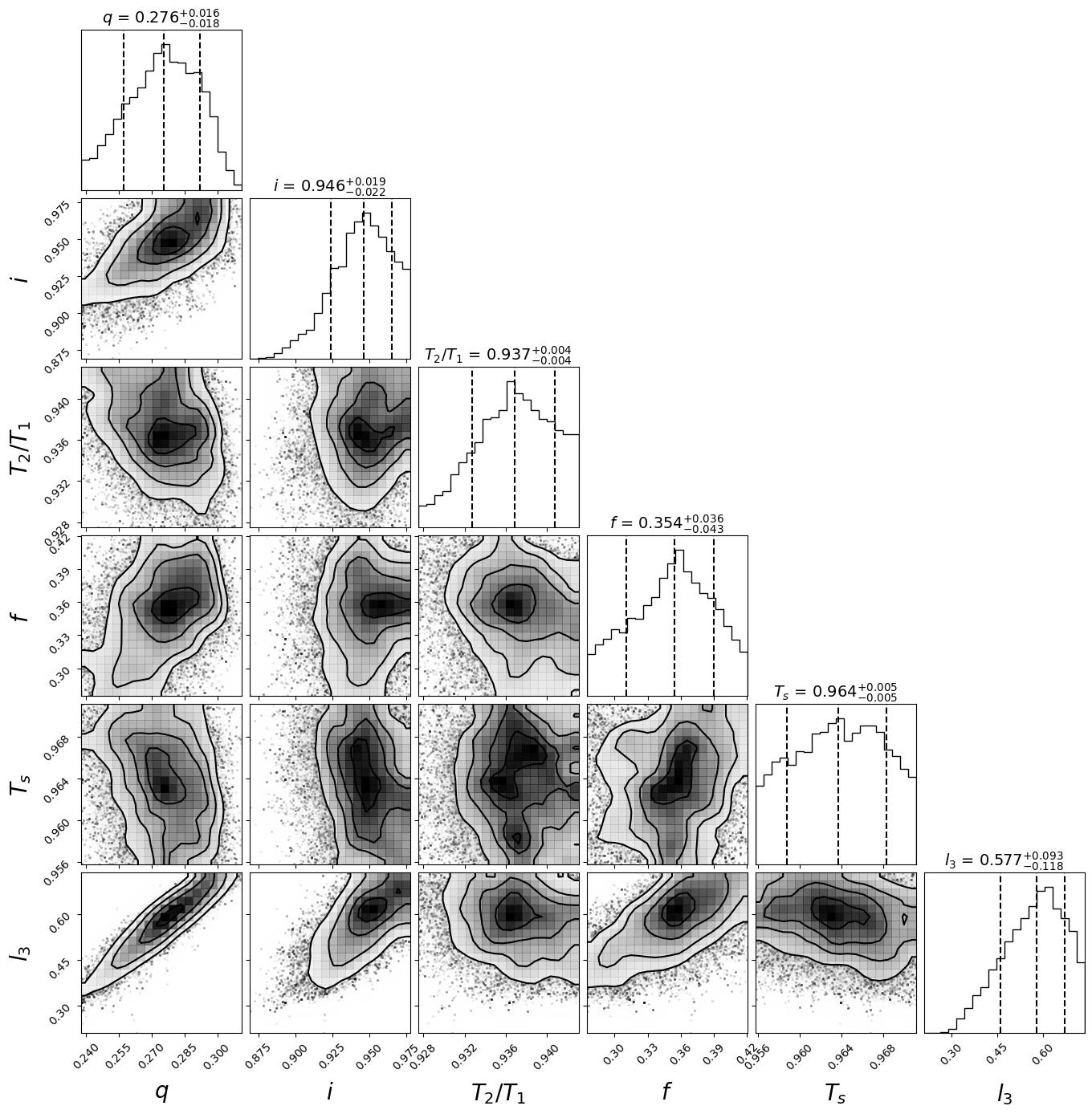}
\includegraphics[width=0.45\textwidth]{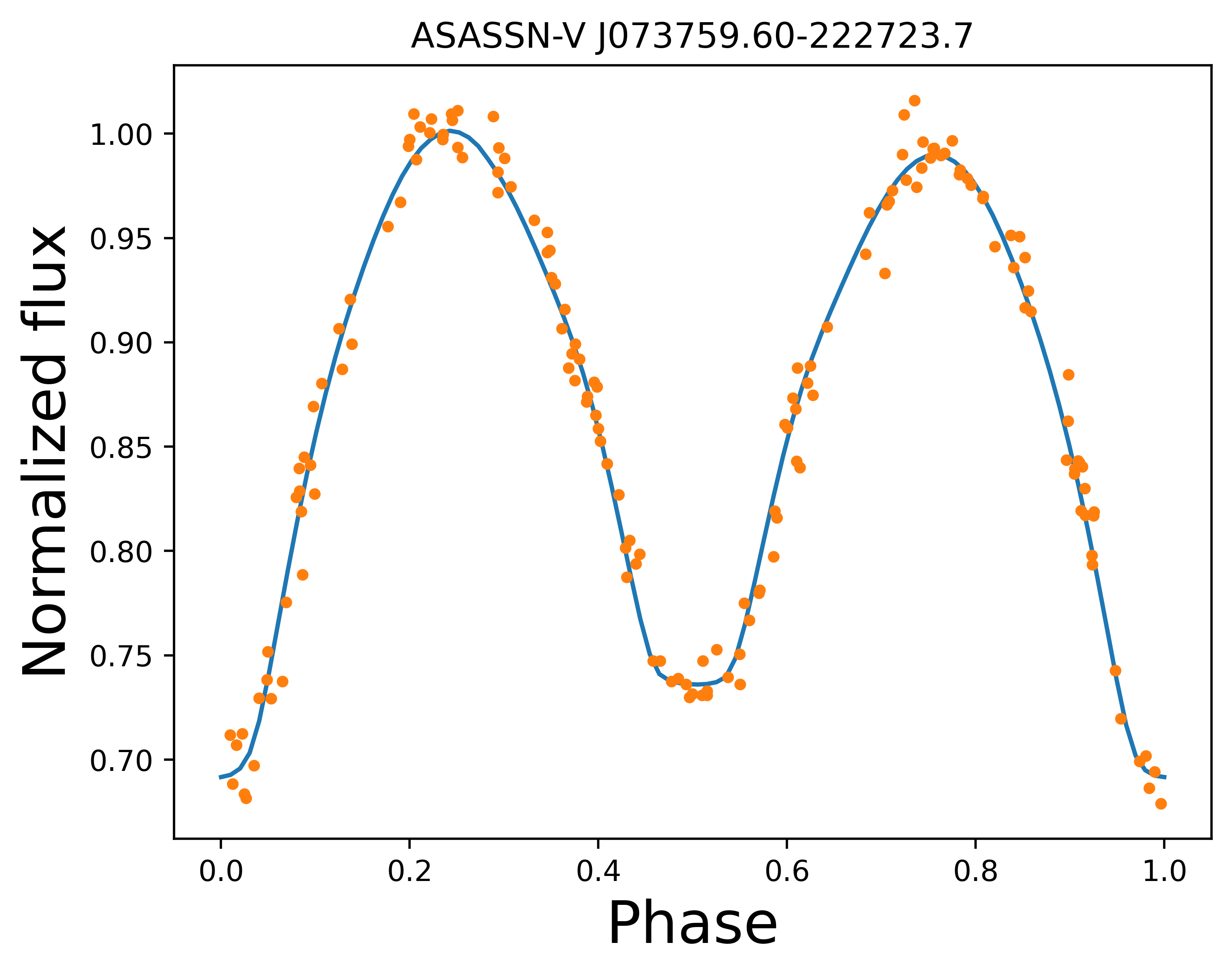}
\caption{The posterior distributions of the parameters and the best fits for two systems.}
\label{Fig2}
\end{figure*}

We calculated the physical parameters using MCMC twice. In the first calculation, 32 walkers and 2000 iterations were employed with the parameters chosen from a uniform distribution over the ranges used in the NN model training. In the second calculation, 16 walkers and 2000 iterations were employed, the results obtained by the first calculation were set as the priors, which follow a Gaussian distribution with 1-$\sigma$ error as the boundary. Since the mass ratios of contact binaries can be greater than 1, we set the phase shift to 0.5 and repeated the two steps loading to four models for each target and the model with the highest R$^2$ was selected as the final model. Systems with R$^2$ less than 0.8 were removed due to the poor fit. Physical parameters for a total of 12201 contact binaries were obtained. Two examples of the best fitting models and the posterior parameter distributions are shown in Figure \ref{Fig2}. 

\section{Statistics of the physical parameters} 
\begin{figure*}
\centering
\includegraphics[width=0.32\textwidth]{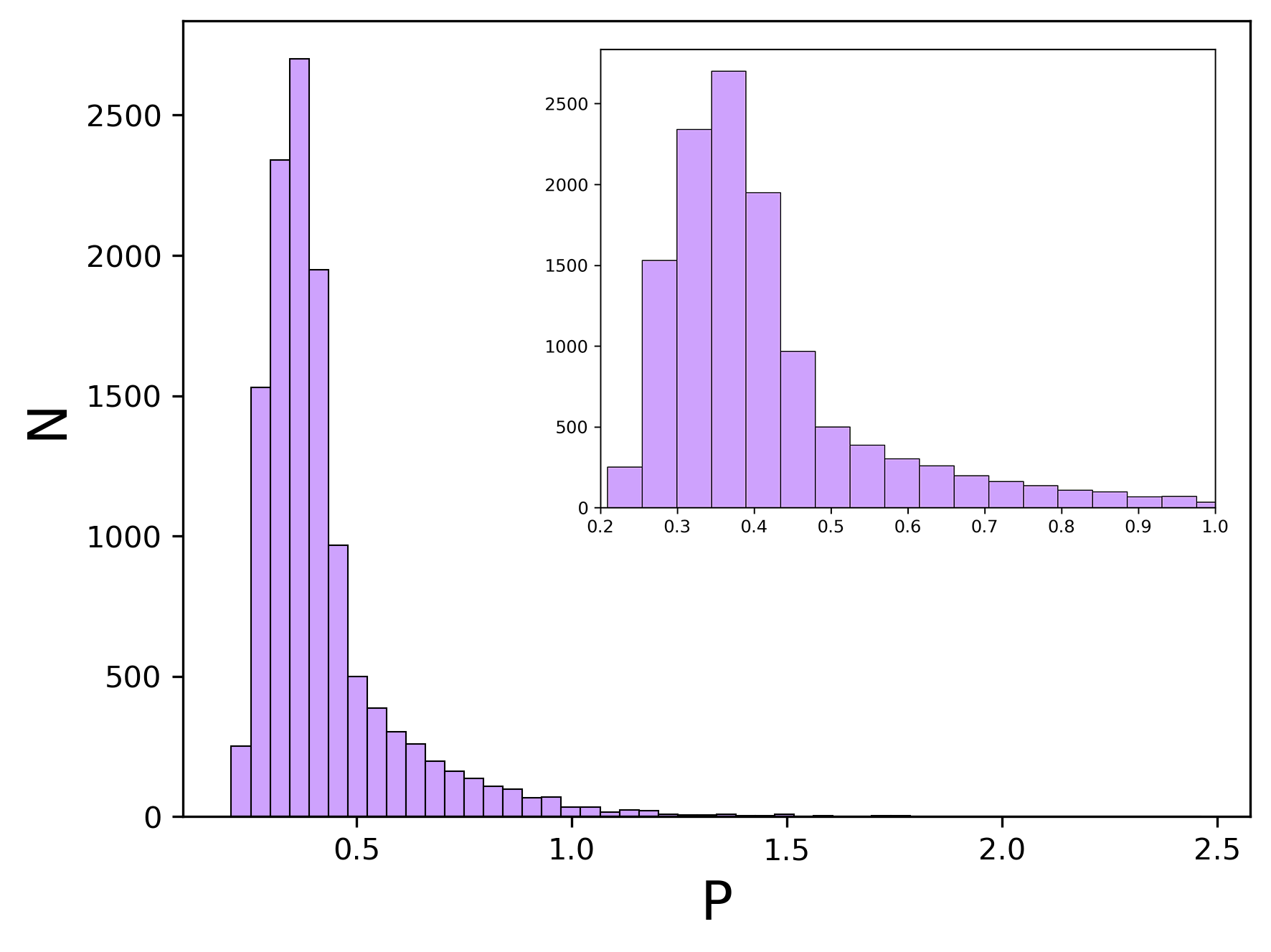}
\includegraphics[width=0.32\textwidth]{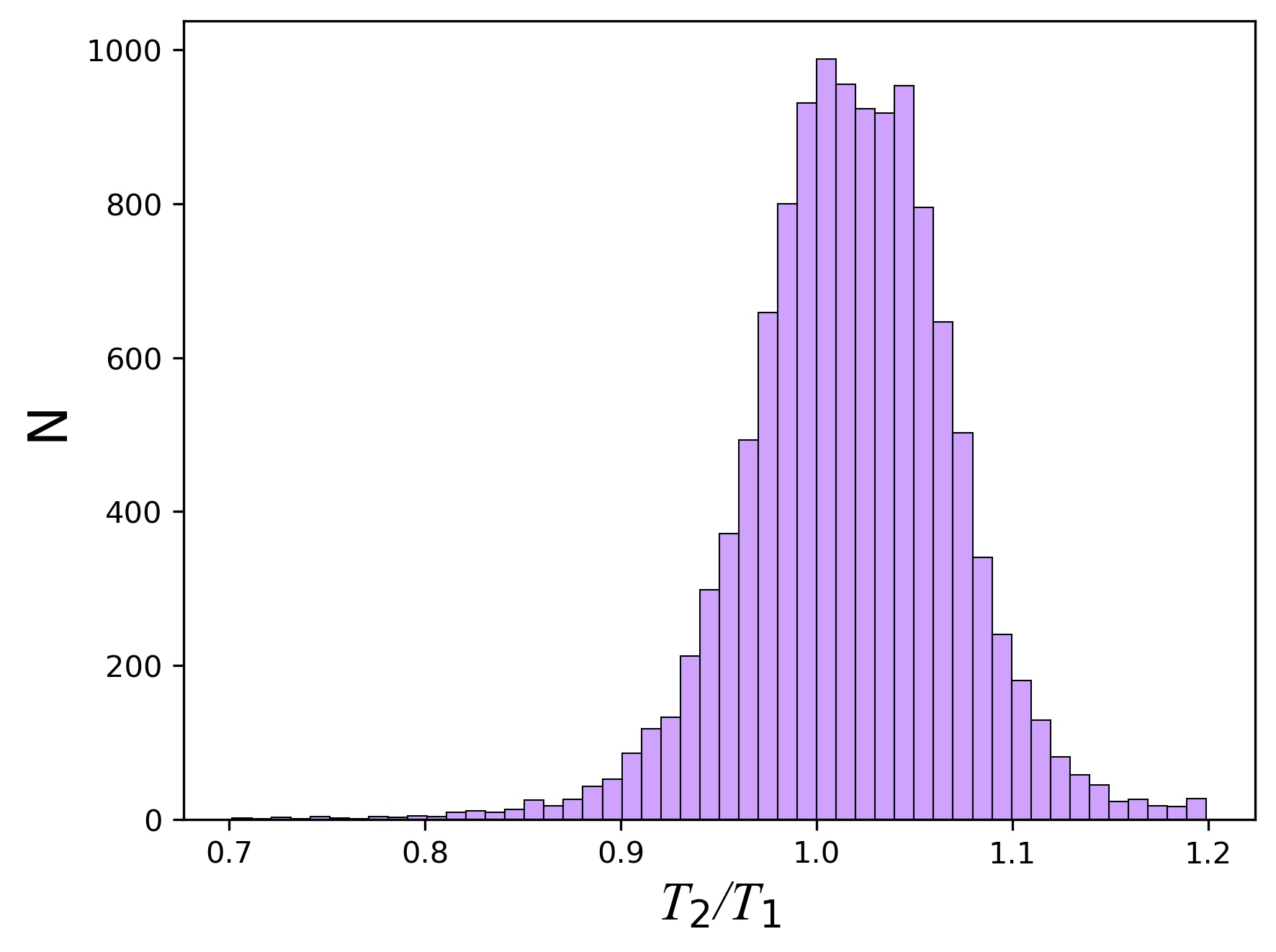}
\includegraphics[width=0.32\textwidth]{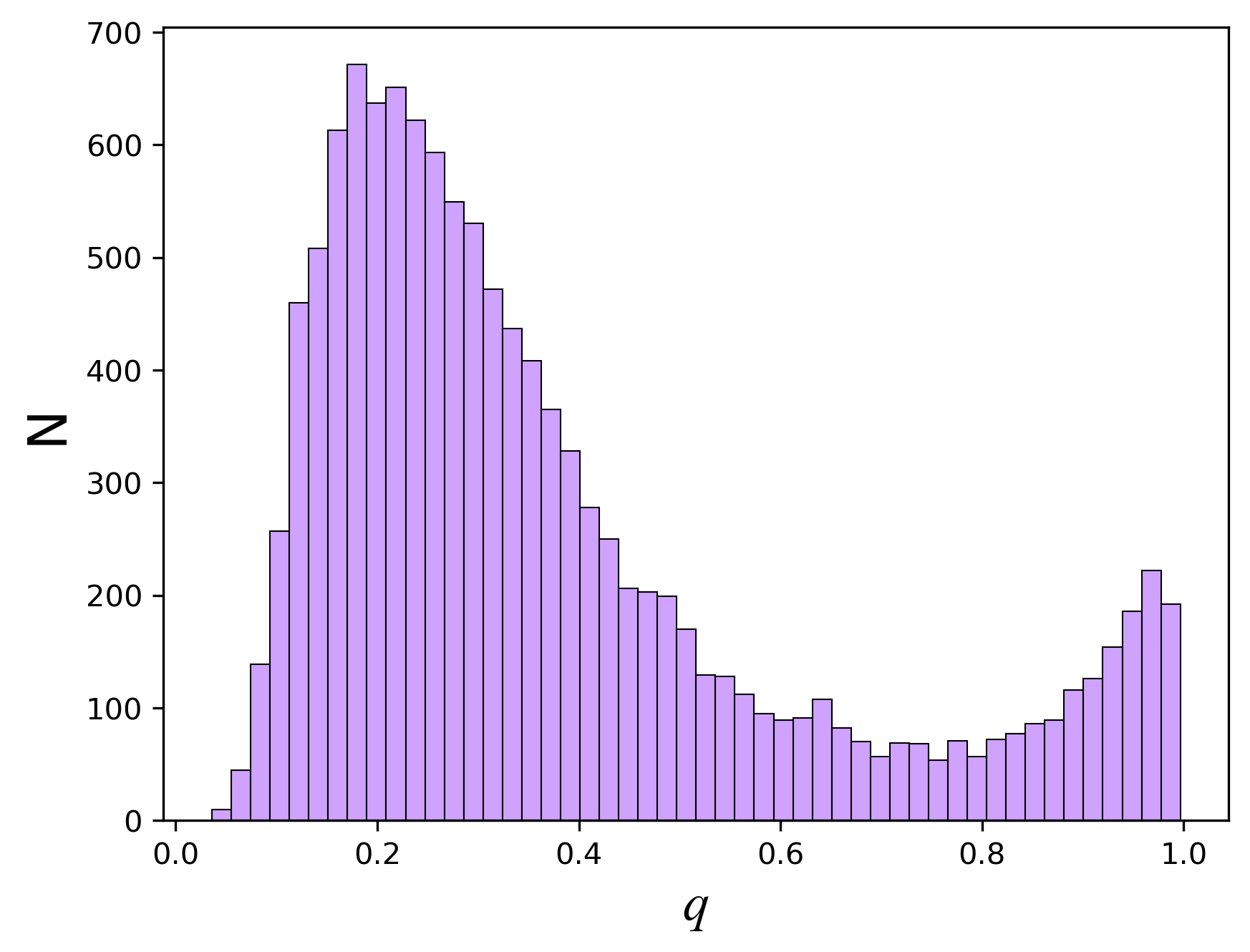}
\includegraphics[width=0.32\textwidth]{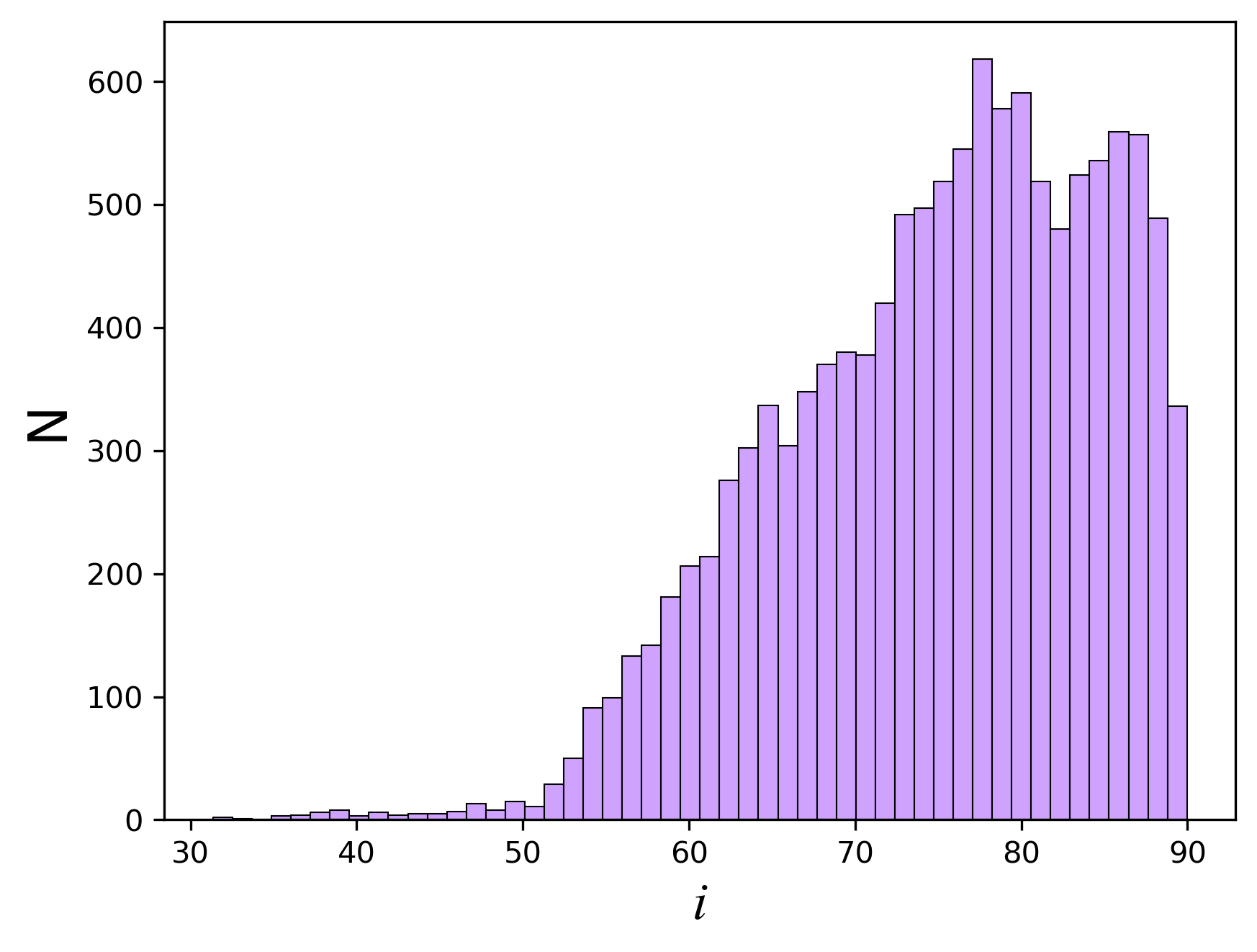}
\includegraphics[width=0.32\textwidth]{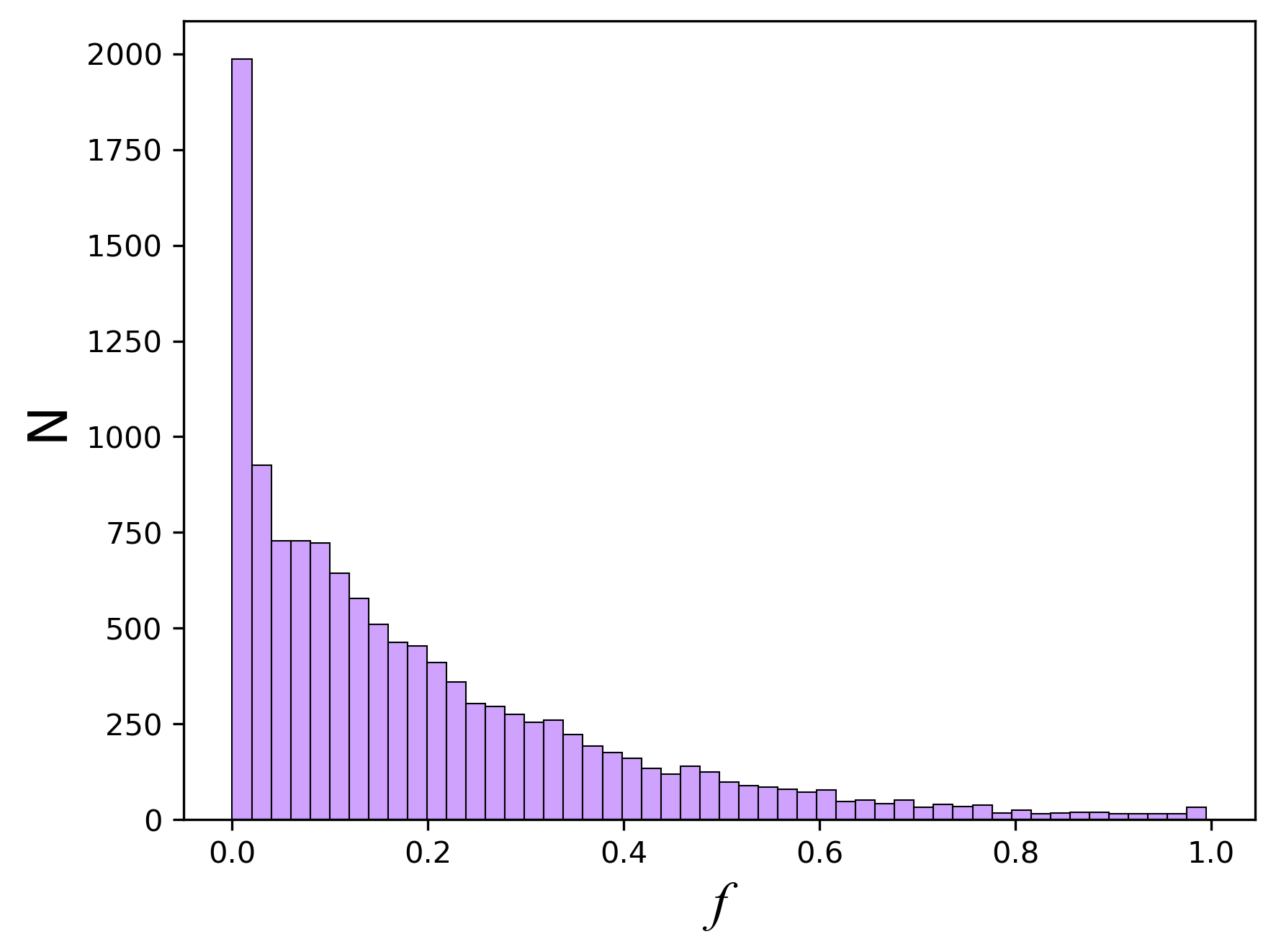}
\includegraphics[width=0.32\textwidth]{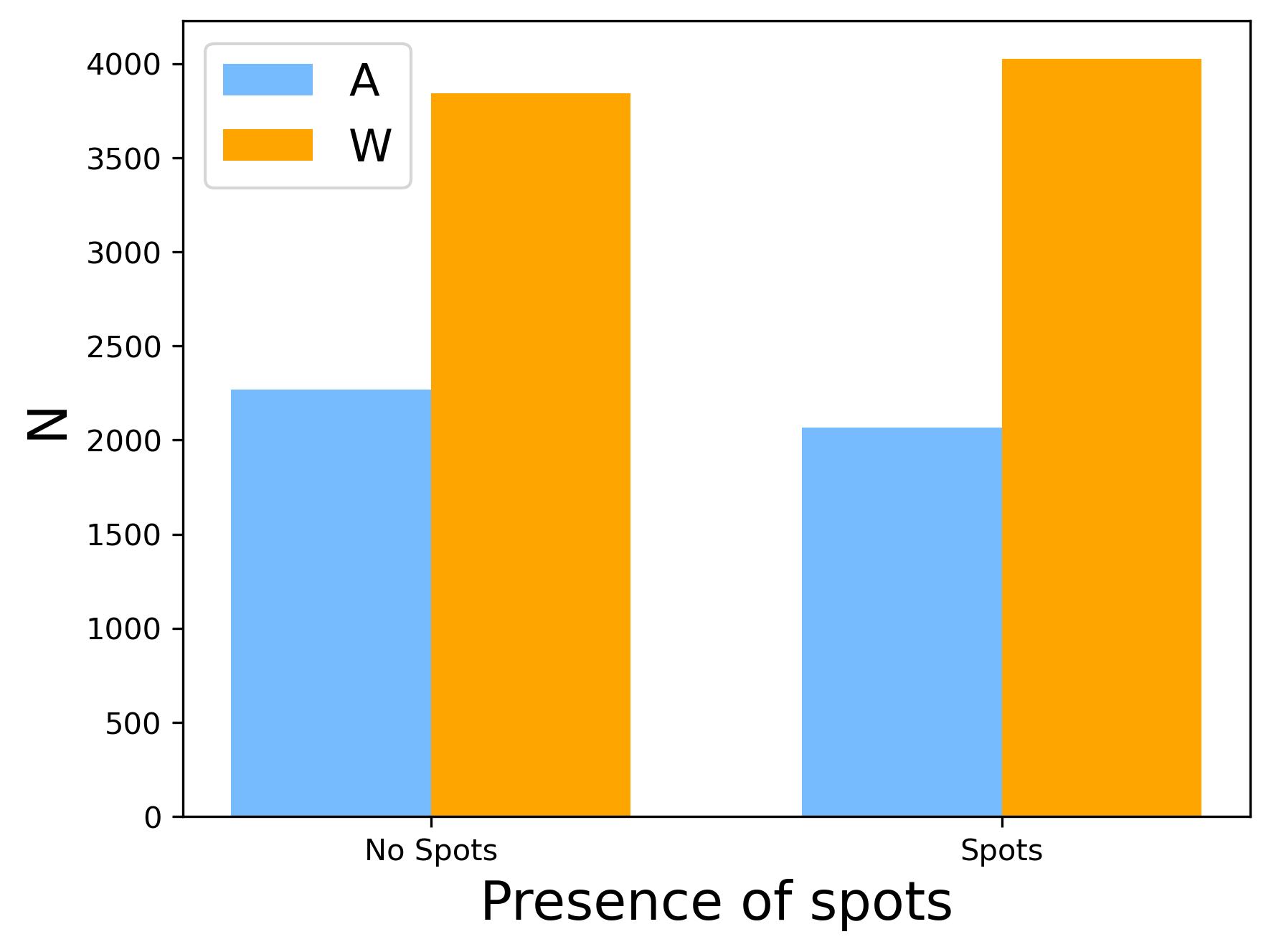}
\caption{The distributions of the binaries in orbital period, $T_2/T_1$, $q$, $i$, $f$, and numbers with spots for the two subtype systems. The inset in the orbital period distribution shows an enlargement of the range from 0.2 to 1 day.}
\label{Fig3}
\end{figure*}

\begin{figure*}
\centering
\includegraphics[width=0.32\textwidth]{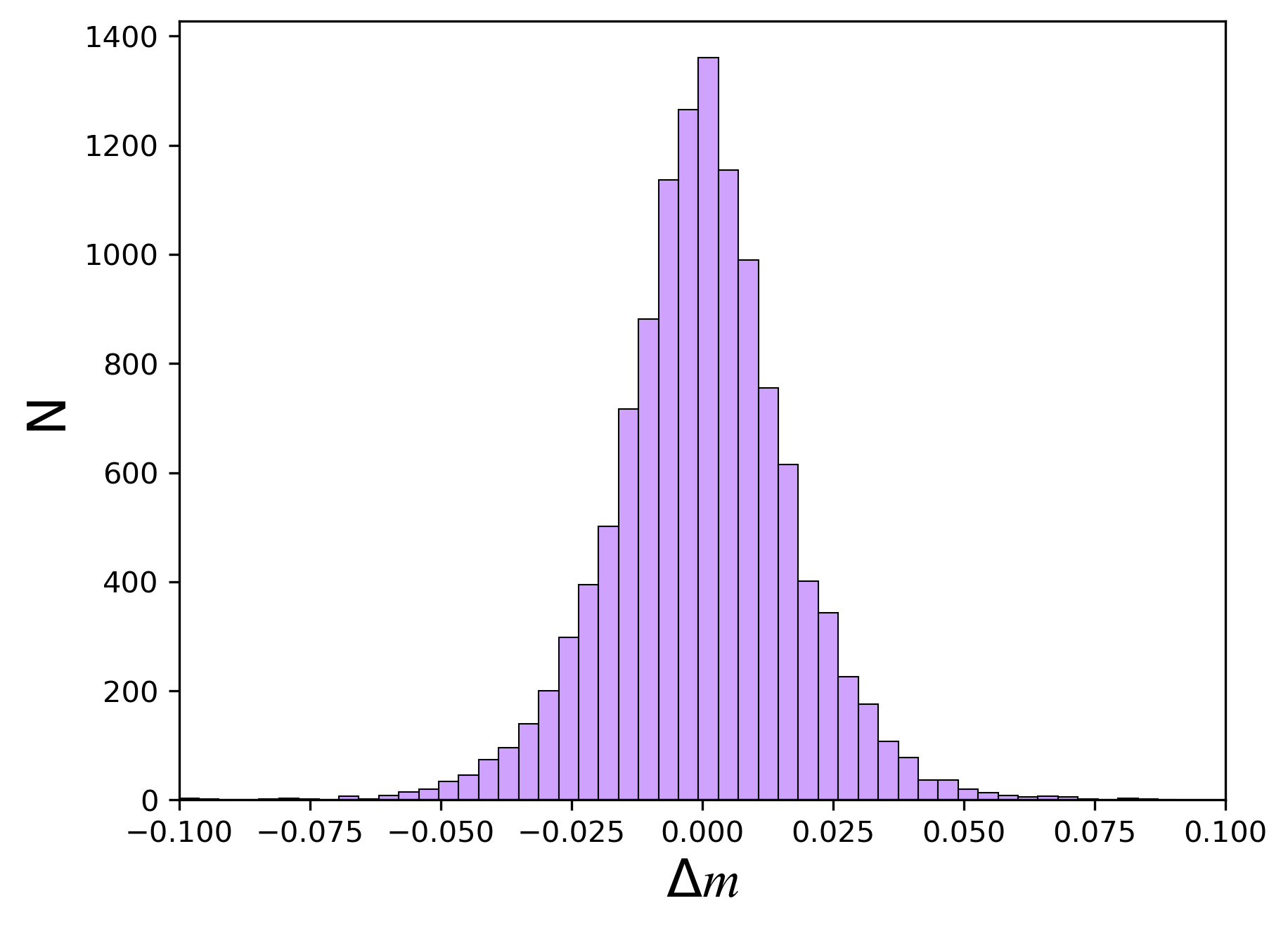}
\includegraphics[width=0.32\textwidth]{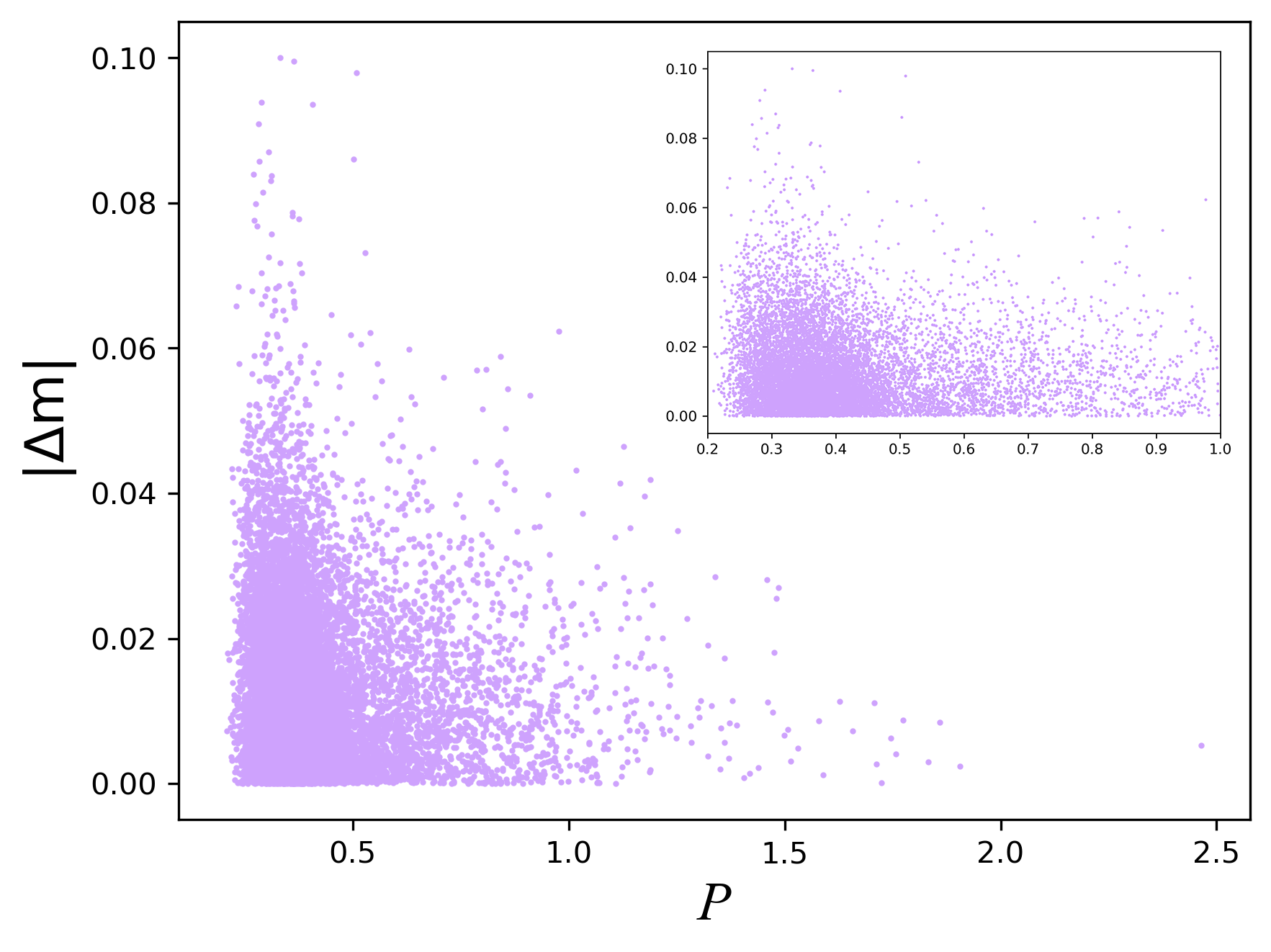}
\includegraphics[width=0.32\textwidth]{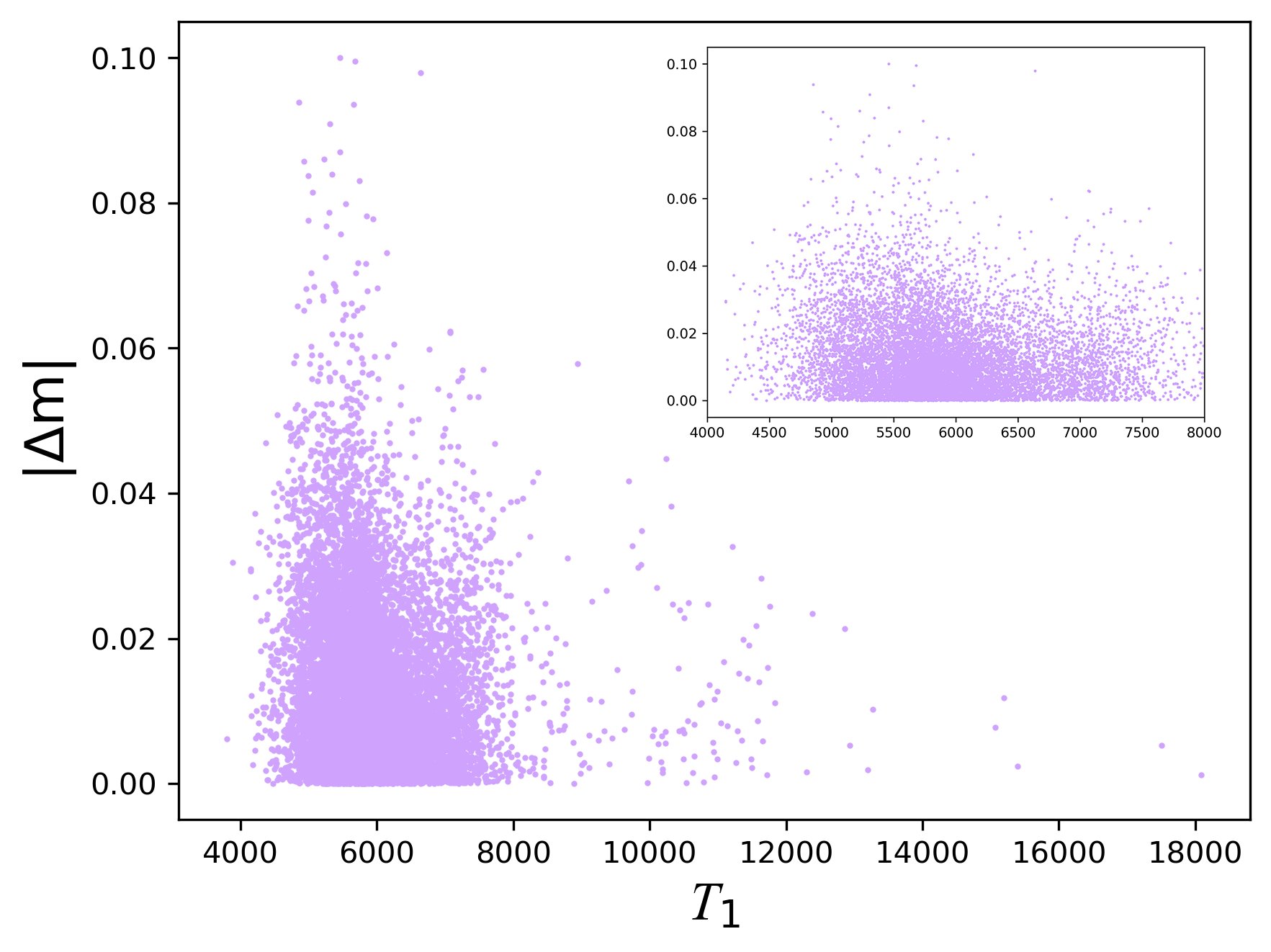}
\caption{The distributions of the differences between the light maxima ($\Delta m$) and the correlations between the orbital period and $\left| \Delta m\right|$, and the temperature of the primary and  $\left| \Delta m\right|$. The inset in the middle figure shows an enlargement of the range from 0.2 to 1 day, and which in the right figure shows an enlargement of the range from 4000 to 8000 K.}
\label{Fig4}
\end{figure*}

 Using Phoebe, we calculated the relative radii of the two components ($r_1$, $r_2$) and the luminosity ratio between the two components ($L_2/L_1$) based on the parameters obtained by NN models. All these physical parameters are listed in Table \ref{tab:physical parameters}.
The distributions of the orbital period, $T_2/T_1$, $q$, $i$, $f$, and number of systems with spots for two subtypes (A- and W- subtypes) are shown in Figure \ref{Fig3}. As seen in this figure, the orbital period distribution has a peak around 0.37 days, $T_2/T_1$ is between 0.8 and 1.2 for most systems, there are two peaks for the mass ratio $q$, one near 0.2, and the other near 0.98, the orbital inclination $i$ is higher than $50^\circ$ for most systems, the fillout factor is less than 60\%  for most systems and nearly half have $f<10\%$, the number of W-subtype systems exceeds that of A-subtype systems, but the probability of spot occurrence for both is around 50\%. In Figure \ref{Fig4}, we plot the distributions of the differences in flux between the light maxima ($\Delta m$), and the correlations between the orbital period and the temperature of the primary component with $\left| \Delta m\right|$. We find that $\left| \Delta m\right|$ is smaller than 0.1 for all systems, that the value of $\left| \Delta m\right|$ decreases as the period and the temperature of the primary component increase, and that G-type contact binaries are more likely to have larger $\left| \Delta m\right|$.

\setlength{\tabcolsep}{1.5pt} 
\begin{table*}
\centering
\flushleft
\fontsize{4.7}{6.5}\selectfont
\caption{Physical parameters of the 12201 ASAS-SN contact binaries} \label{tab:physical parameters}
\begin{tabular}{lccccclccccccccccccclccccc}
\hline
Name (ASASSN-V+)    & RA         & DEC          &P(d)          &Min.I      & error        &Type        &$T_1$ (K)&$q$      &error    &$i$        &error    &$T_2/T_1$&error    &$f    $   &error    &$l_3$    &error    &$T_s$    &error    &long    &$L_2/L_1$    &$r_1$    &$r_2$    &$\Omega$    &$R^2$\\
\hline
J000030.35+391106.8    & 0.126  &    39.185      &0.28799      &2457997.2629  &    0.0004      &W          &5655       &0.255     &0.010     &78.745     &0.718     &1.072       &0.002     &0.186     &0.015     &0.533     &0.045     &0.000     &0.000     &0      &0.393         &0.511     &0.279     &2.335       &0.951\\ 
J000052.76+370428.1    & 0.220  &    37.074      &0.41756      &2457306.3106  &    0.0005      &A          &6165       &0.528     &0.062     &64.002     &1.067     &0.969       &0.003     &0.274     &0.047     &0.378     &0.117     &0.000     &0.000     &0      &0.505         &0.457     &0.347     &2.843       &0.974\\ 
J000057.61+023641.3    & 0.240  &    2.611       &0.31808      &2457981.3098  &    0.0007      &W          &6013       &0.256     &0.010     &86.788     &1.664     &1.022       &0.004     &0.296     &0.027     &0.361     &0.038     &0.000     &0.000     &0      &0.333         &0.517     &0.285     &2.318       &0.962\\ 
J000107.66-333003.8    & 0.282  &    -33.501     &0.46657      &2458330.4603  &    0.0008      &H/W        &5941       &0.926     &0.079     &67.743     &0.121     &1.040       &0.002     &0.158     &0.009     &0.000     &0.000     &0.943     &0.006     &270    &1.091         &0.401     &0.388     &3.548       &0.937\\ 
J000145.82+712002.9    & 0.441  &    71.334      &0.69035      &2458098.1702  &    0.0008      &W          &7335       &0.332     &0.096     &69.325     &0.751     &1.028       &0.056     &0.169     &0.028     &0.000     &0.000     &0.996     &0.003     &90      &0.418         &0.488     &0.299     &2.502       &0.967\\ 
J000150.41+332253.8    & 0.460  &    33.382      &0.29909      &2457360.1407  &    0.0007      &H/A        &5508       &0.849     &0.086     &73.365     &0.677     &0.974       &0.003     &0.145     &0.021     &0.502     &0.067     &0.942     &0.005     &270    &0.777         &0.408     &0.379     &3.431       &0.910\\ 
J000201.67-665319.1    & 0.507  &    -66.889     &0.32658      &2458361.0556  &    0.0003      &W          &5741       &0.252     &0.008     &81.590     &0.608     &1.041       &0.002     &0.107     &0.013     &0.040     &0.027     &0.000     &0.000     &0      &0.340         &0.508     &0.273     &2.341       &0.968\\ 
J000248.22+151151.6    & 0.701  &    15.198      &0.38097      &2457679.0389  &    0.0008      &W          &6215       &0.248     &0.005     &87.646     &1.569     &1.007       &0.002     &0.430     &0.025     &0.499     &0.021     &0.999     &0.001     &90      &0.314         &0.526     &0.291     &2.281       &0.889\\ 
J000253.82+671301.6    & 0.724  &    67.217      &0.35913      &2458349.4308  &    0.0005      &A          &5866       &0.305     &0.018     &73.930     &0.586     &0.995       &0.005     &0.006     &0.006     &0.576     &0.021     &0.000     &0.000     &0      &0.328         &0.487     &0.282     &2.476       &0.858\\ 
J000309.29-345618.1    & 0.789  &    -34.938     &0.26542      &2458085.9253  &    0.0004      &W          &4823       &0.335     &0.014     &75.221     &0.630     &1.070       &0.002     &0.005     &0.005     &0.484     &0.057     &0.000     &0.000     &0      &0.479         &0.478     &0.289     &2.541       &0.906\\ 
\hline
\end{tabular}
\vspace{6pt} 
\footnotesize
\begin{tablenotes}[para]
\item (This table is available in its entirety in machine-readable form.)\\
\end{tablenotes}
\end{table*}

\begin{figure*}
\centering
\includegraphics[width=0.47\textwidth]{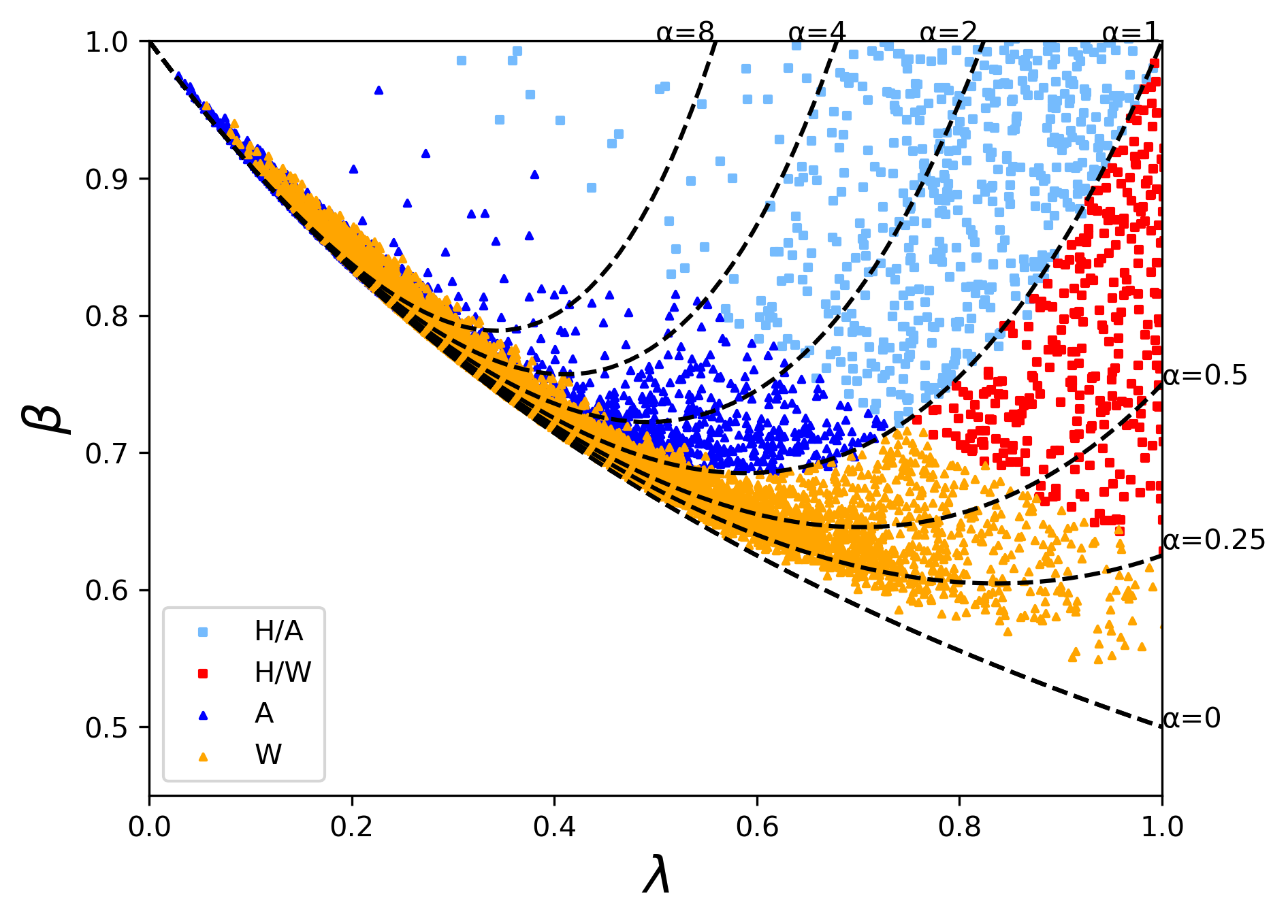}
\includegraphics[width=0.45\textwidth]{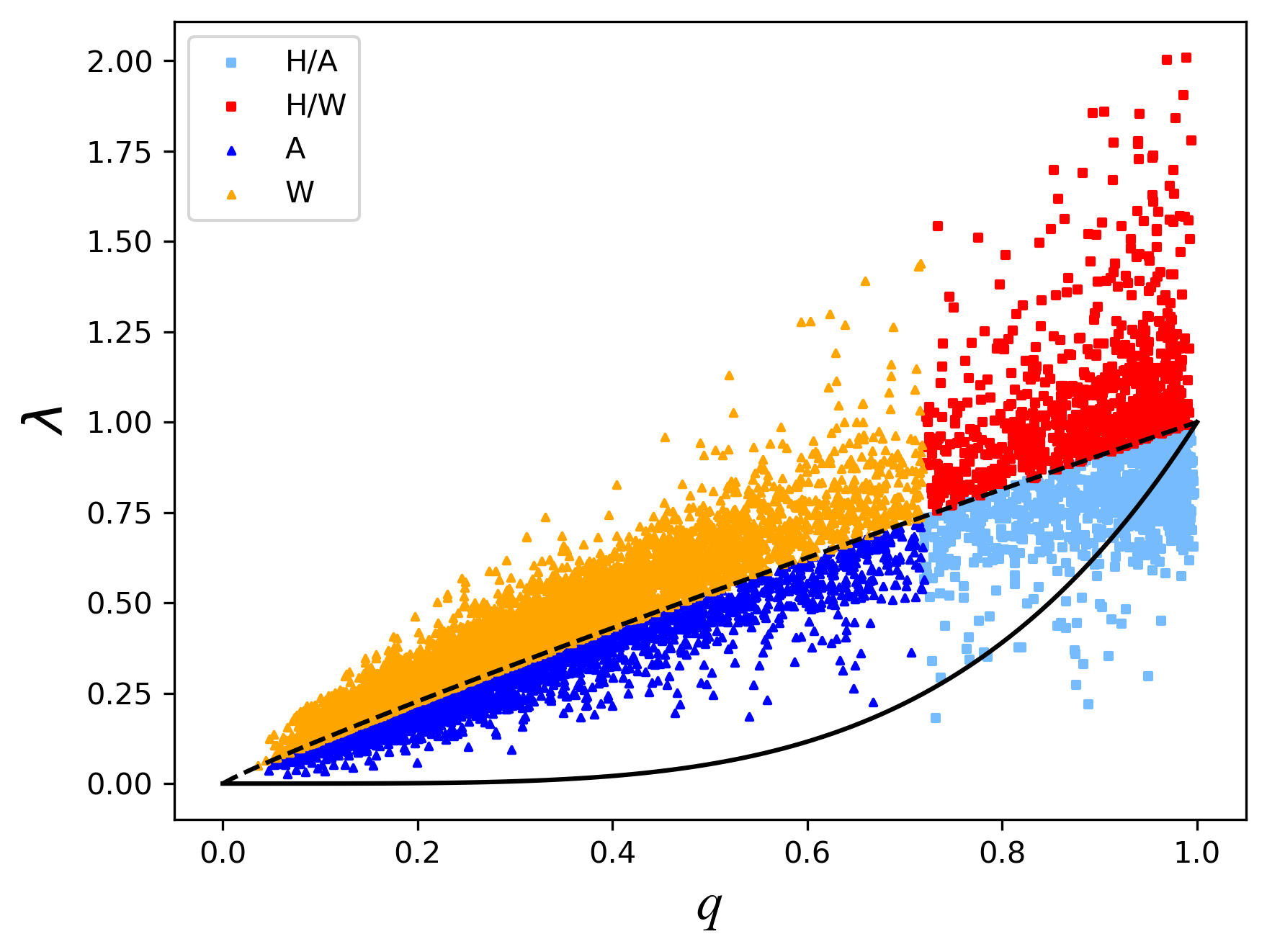}
\caption{The left panel shows the distribution in transfer parameter $\beta$ and luminosity ratio $\lambda$. Dashed lines are the expected $\beta$ curves for different values of $\alpha$. The right panel shows the relationship between bolometric luminosity ratio and mass ratio ($q$). The dashed line is from \cite{1968ApJ...153..877L}'s relation $\lambda=q^{0.92}$, and the solid line is the main-sequence mass luminosity relation, $\lambda=q^{4.216}$.}
\label{Fig5}
\end{figure*}

\cite{2004A&A...426.1001C} proposed the definitions of H-subtype contact binaries ($q>0.72$) and the transfer parameter,
\begin{equation}
\beta=\frac{L_{1,observed}}{L_{1,ZAMS}}=\frac{1+q^{4.216}}{1+q^{0.92}\left(\frac{T_{2}}{T_{1}}\right)^{4}}=\frac{1+\alpha\lambda^{4.58}}{1+\lambda},
\end{equation}
where $L_{1,observed}$ is the observed luminosity of the primary component, $L_{1,ZAMS}$ is the luminosity of the primary component at zero age main sequence (ZAMS), $\alpha=(\frac{T_1}{T_2})^{18.3}$, and $\lambda$ is the bolometric luminosity ratio (\citealt{2020ApJS..247...50S} corrected the indices of $\alpha$ and $\lambda$). We calculated the transfer parameters for all systems, and divided them into four groups (A-subtype systems with $q\leq0.72$, A-subtype systems with $q>0.72$, W-subtype systems with $q\leq0.72$, and W-subtype systems with $q>0.72$) and plot them in the left panel of Figure \ref{Fig5}. As shown in Figure \ref{Fig5}, the four types of system are distributed in different regions: A-subtype contact binaries have higher transfer parameters than the W-subtype, and H-subtype systems exhibit higher transfer parameters, indicating less efficient energy transfer rate than the other types. The right panel of Figure \ref{Fig5} shows the relationship between bolometric luminosity ratio and mass ratio ($q$). \cite{1968ApJ...153..877L} found a relation between bolometric luminosity ratio and mass ratio ($\lambda=q^{0.92}$). We find that A- and W-subtype systems are roughly divided by this relationship and that some A-subtype systems with $q>0.72$ are below the main-sequence mass luminosity relation, indicating that no or less mass transfer between the two components.

\section{Conclusions} 
This paper constructed NN models trained on Phoebe generated LCs to quickly determine the physical parameters of contact binaries. Because a large number of contact binaries exhibit asymmetric LCs, we included star spots in our models. 12785 ASAS-SN contact binaries were selected by two criteria and analyzed by our NN models and the MCMC algorithm. The physical parameters of 12201 contact binaries with fit statistics $R^2>0.8$ are obtained. Among these binaries, 4332 are A-subtype systems, and 7869 are W-type systems. We examined the distributions in the orbital period, $T_2/T_1$, $q$, $i$, $f$, the probability of the presence of spots, distributions of $\Delta m$, and the correlations between the orbital period, the temperature of the primary component with $\left| \Delta m\right|$.

Comparing to the studies of \cite{2023MNRAS.525.4596D}, \cite{2024ApJS..271...32L}, and \cite{2024ApJS..273...31W}, we have introduced spot parameters, which enhance the precision of the photometric solutions for contact binaries through machine learning method. However, only one spot parameter (the relative temperature $T_s$) was set adjustable and only one filter was used in our models, future studies should focus on developing models where all four spot parameters are adjustable and that can be applied to multi-band LCs.

\section{acknowledgments}
We are deeply grateful to the referee for the insightful comments and constructive suggestions, which have significantly enhanced the quality of this manuscript. This work is supported by National Natural Science Foundation of China (NSFC) (No. 12273018), and the Joint Research Fund in Astronomy (No. U1931103) under cooperative agreement between NSFC and Chinese Academy of Sciences (CAS), and by the Qilu Young Researcher Project of Shandong University, and by the Young Data Scientist Project of the National Astronomical Data Center, and by the Cultivation Project for LAMOST Scientific Payoff and Research Achievement of CAMS-CAS. The calculations in this work were carried out at Supercomputing Center of Shandong University, Weihai.

This paper makes use of data from ASAS-SN. ASAS-SN is funded
in part by the Gordon and Betty Moore Foundation through grant
numbers GBMF5490 and GBMF10501 to the Ohio State University,
and also funded in part by the Alfred P. Sloan Foundation grant
number G-2021-14192.

This work has made use of data from the European Space Agency (ESA) mission
{\it Gaia} (\url{https://www.cosmos.esa.int/gaia}), processed by the {\it Gaia}
Data Processing and Analysis Consortium (DPAC,
\url{https://www.cosmos.esa.int/web/gaia/dpac/consortium}). Funding for the DPAC
has been provided by national institutions, in particular the institutions
participating in the {\it Gaia} Multilateral Agreement.

\bibliography{sample631}{}

\begin{thebibliography}{}
\expandafter\ifx\csname natexlab\endcsname\relax\def\natexlab#1{#1}\fi
\providecommand{\url}[1]{\href{#1}{#1}}
\providecommand{\dodoi}[1]{doi:~\href{http://doi.org/#1}{\nolinkurl{#1}}}
\providecommand{\doeprint}[1]{\href{http://ascl.net/#1}{\nolinkurl{http://ascl.net/#1}}}
\providecommand{\doarXiv}[1]{\href{https://arxiv.org/abs/#1}{\nolinkurl{https://arxiv.org/abs/#1}}}

\bibitem[{{Arbutina} \& {Wadhwa}(2024)}]{2024SerAJ.208....1A}
{Arbutina}, B., \& {Wadhwa}, S. 2024, Serbian Astronomical Journal, 208, 1

\bibitem[{{Bradstreet} \& {Guinan}(1994)}]{1994ASPC...56..228B}
{Bradstreet}, D.~H., \& {Guinan}, E.~F. 1994, in Astronomical Society of the
  Pacific Conference Series, Vol.~56, Interacting Binary Stars, ed. A.~W.
  {Shafter}, 228

\bibitem[{{Castelli} \& {Kurucz}(2004)}]{2004A&A...419..725C}
{Castelli}, F., \& {Kurucz}, R.~L. 2004, \aap, 419, 725

\bibitem[{{Chen} {et~al.}(2016){Chen}, {de Grijs}, \&
  {Deng}}]{2016ApJ...832..138C}
{Chen}, X., {de Grijs}, R., \& {Deng}, L. 2016, \apj, 832, 138

\bibitem[{{Chen} {et~al.}(2018){Chen}, {Deng}, {de Grijs}, {Wang}, \&
  {Feng}}]{2018ApJ...859..140C}
{Chen}, X., {Deng}, L., {de Grijs}, R., {Wang}, S., \& {Feng}, Y. 2018, \apj,
  859, 140

\bibitem[{{Chen} {et~al.}(2020){Chen}, {Wang}, {Deng}, {de Grijs}, {Yang}, \&
  {Tian}}]{2020ApJS..249...18C}
{Chen}, X., {Wang}, S., {Deng}, L., {et~al.} 2020, \apjs, 249, 18

\bibitem[{{Clarke}(2002)}]{2002A&A...386..763C}
{Clarke}, D. 2002, \aap, 386, 763, \dodoi{10.1051/0004-6361:20020258}

\bibitem[{{Csizmadia} \& {Klagyivik}(2004)}]{2004A&A...426.1001C}
{Csizmadia}, S., \& {Klagyivik}, P. 2004, \aap, 426, 1001,
  \dodoi{10.1051/0004-6361:20040430}

\bibitem[{{Ding} {et~al.}(2023){Ding}, {Ji}, {Li}, {Xiong}, {Cheng}, \&
  {Wang}}]{2023MNRAS.525.4596D}
{Ding}, X., {Ji}, K., {Li}, X., {et~al.} 2023, \mnras, 525, 4596,
  \dodoi{10.1093/mnras/stad2565}

\bibitem[{{Ding} {et~al.}(2022){Ding}, {Ji}, {Li}, {Xiong}, {Cheng}, {Wang}, \&
  {Liu}}]{2022AJ....164..200D}
---. 2022, \aj, 164, 200, \dodoi{10.3847/1538-3881/ac8e66}

\bibitem[{{Foreman-Mackey} {et~al.}(2019){Foreman-Mackey}, {Farr}, {Sinha},
  {Archibald}, {Hogg}, {Sanders}, {Zuntz}, {Williams}, {Nelson}, {de
  Val-Borro}, {Erhardt}, {Pashchenko}, \& {Pla}}]{2019JOSS....4.1864F}
{Foreman-Mackey}, D., {Farr}, W., {Sinha}, M., {et~al.} 2019, The Journal of
  Open Source Software, 4, 1864, \dodoi{10.21105/joss.01864}

\bibitem[{{Gaia Collaboration} {et~al.}(2016){Gaia Collaboration}, {Prusti},
  {de Bruijne}, {Brown}, {Vallenari}, {Babusiaux}, {Bailer-Jones}, {Bastian},
  {Biermann}, {Evans}, {Eyer}, {Jansen}, {Jordi}, {Klioner}, {Lammers},
  {Lindegren}, {Luri}, {Mignard}, {Milligan}, {Panem}, {Poinsignon},
  {Pourbaix}, {Randich}, {Sarri}, {Sartoretti}, {Siddiqui}, {Soubiran},
  {Valette}, {van Leeuwen}, {Walton}, {Aerts}, {Arenou}, {Cropper}, {Drimmel},
  {H{\o}g}, {Katz}, {Lattanzi}, {O'Mullane}, {Grebel}, {Holland}, {Huc},
  {Passot}, {Bramante}, {Cacciari}, {Casta{\~n}eda}, {Chaoul}, {Cheek}, {De
  Angeli}, {Fabricius}, {Guerra}, {Hern{\'a}ndez}, {Jean-Antoine-Piccolo},
  {Masana}, {Messineo}, {Mowlavi}, {Nienartowicz}, {Ord{\'o}{\~n}ez-Blanco},
  {Panuzzo}, {Portell}, {Richards}, {Riello}, {Seabroke}, {Tanga},
  {Th{\'e}venin}, {Torra}, {Els}, {Gracia-Abril}, {Comoretto},
  {Garcia-Reinaldos}, {Lock}, {Mercier}, {Altmann}, {Andrae}, {Astraatmadja},
  {Bellas-Velidis}, {Benson}, {Berthier}, {Blomme}, {Busso}, {Carry},
  {Cellino}, {Clementini}, {Cowell}, {Creevey}, {Cuypers}, {Davidson}, {De
  Ridder}, {de Torres}, {Delchambre}, {Dell'Oro}, {Ducourant}, {Fr{\'e}mat},
  {Garc{\'\i}a-Torres}, {Gosset}, {Halbwachs}, {Hambly}, {Harrison}, {Hauser},
  {Hestroffer}, {Hodgkin}, {Huckle}, {Hutton}, {Jasniewicz}, {Jordan},
  {Kontizas}, {Korn}, {Lanzafame}, {Manteiga}, {Moitinho}, {Muinonen},
  {Osinde}, {Pancino}, {Pauwels}, {Petit}, {Recio-Blanco}, {Robin}, {Sarro},
  {Siopis}, {Smith}, {Smith}, {Sozzetti}, {Thuillot}, {van Reeven}, {Viala},
  {Abbas}, {Abreu Aramburu}, {Accart}, {Aguado}, {Allan}, {Allasia},
  {Altavilla}, {{\'A}lvarez}, {Alves}, {Anderson}, {Andrei}, {Anglada Varela},
  {Antiche}, {Antoja}, {Ant{\'o}n}, {Arcay}, {Atzei}, {Ayache}, {Bach},
  {Baker}, {Balaguer-N{\'u}{\~n}ez}, {Barache}, {Barata}, {Barbier}, {Barblan},
  {Baroni}, {Barrado y Navascu{\'e}s}, {Barros}, {Barstow}, {Becciani},
  {Bellazzini}, {Bellei}, {Bello Garc{\'\i}a}, {Belokurov}, {Bendjoya},
  {Berihuete}, {Bianchi}, {Bienaym{\'e}}, {Billebaud}, {Blagorodnova},
  {Blanco-Cuaresma}, {Boch}, {Bombrun}, {Borrachero}, {Bouquillon}, {Bourda},
  {Bouy}, {Bragaglia}, {Breddels}, {Brouillet}, {Br{\"u}semeister},
  {Bucciarelli}, {Budnik}, {Burgess}, {Burgon}, {Burlacu}, {Busonero}, {Buzzi},
  {Caffau}, {Cambras}, {Campbell}, {Cancelliere}, {Cantat-Gaudin}, {Carlucci},
  {Carrasco}, {Castellani}, {Charlot}, {Charnas}, {Charvet}, {Chassat},
  {Chiavassa}, {Clotet}, {Cocozza}, {Collins}, {Collins}, {Costigan}, {Crifo},
  {Cross}, {Crosta}, {Crowley}, {Dafonte}, {Damerdji}, {Dapergolas}, {David},
  {David}, {De Cat}, {de Felice}, {de Laverny}, {De Luise}, {De March}, {de
  Martino}, {de Souza}, {Debosscher}, {del Pozo}, {Delbo}, {Delgado},
  {Delgado}, {di Marco}, {Di Matteo}, {Diakite}, {Distefano}, {Dolding}, {Dos
  Anjos}, {Drazinos}, {Dur{\'a}n}, {Dzigan}, {Ecale}, {Edvardsson}, {Enke},
  {Erdmann}, {Escolar}, {Espina}, {Evans}, {Eynard Bontemps}, {Fabre},
  {Fabrizio}, {Faigler}, {Falc{\~a}o}, {Farr{\`a}s Casas}, {Faye}, {Federici},
  {Fedorets}, {Fern{\'a}ndez-Hern{\'a}ndez}, {Fernique}, {Fienga}, {Figueras},
  {Filippi}, {Findeisen}, {Fonti}, {Fouesneau}, {Fraile}, {Fraser}, {Fuchs},
  {Furnell}, {Gai}, {Galleti}, {Galluccio}, {Garabato}, {Garc{\'\i}a-Sedano},
  {Gar{\'e}}, {Garofalo}, {Garralda}, {Gavras}, {Gerssen}, {Geyer}, {Gilmore},
  {Girona}, {Giuffrida}, {Gomes}, {Gonz{\'a}lez-Marcos},
  {Gonz{\'a}lez-N{\'u}{\~n}ez}, {Gonz{\'a}lez-Vidal}, {Granvik}, {Guerrier},
  {Guillout}, {Guiraud}, {G{\'u}rpide}, {Guti{\'e}rrez-S{\'a}nchez}, {Guy},
  {Haigron}, {Hatzidimitriou}, {Haywood}, {Heiter}, {Helmi}, {Hobbs},
  {Hofmann}, {Holl}, {Holland}, {Hunt}, {Hypki}, {Icardi}, {Irwin}, {Jevardat
  de Fombelle}, {Jofr{\'e}}, {Jonker}, {Jorissen}, {Julbe}, {Karampelas},
  {Kochoska}, {Kohley}, {Kolenberg}, {Kontizas}, {Koposov}, {Kordopatis},
  {Koubsky}, {Kowalczyk}, {Krone-Martins}, {Kudryashova}, {Kull}, {Bachchan},
  {Lacoste-Seris}, {Lanza}, {Lavigne}, {Le Poncin-Lafitte}, {Lebreton},
  {Lebzelter}, {Leccia}, {Leclerc}, {Lecoeur-Taibi}, {Lemaitre}, {Lenhardt},
  {Leroux}, {Liao}, {Licata}, {Lindstr{\o}m}, {Lister}, {Livanou}, {Lobel},
  {L{\"o}ffler}, {L{\'o}pez}, {Lopez-Lozano}, {Lorenz}, {Loureiro},
  {MacDonald}, {Magalh{\~a}es Fernandes}, {Managau}, {Mann}, {Mantelet},
  {Marchal}, {Marchant}, {Marconi}, {Marie}, {Marinoni}, {Marrese},
  {Marschalk{\'o}}, {Marshall}, {Mart{\'\i}n-Fleitas}, {Martino}, {Mary},
  {Matijevi{\v{c}}}, {Mazeh}, {McMillan}, {Messina}, {Mestre}, {Michalik},
  {Millar}, {Miranda}, {Molina}, {Molinaro}, {Molinaro}, {Moln{\'a}r},
  {Moniez}, {Montegriffo}, {Monteiro}, {Mor}, {Mora}, {Morbidelli}, {Morel},
  {Morgenthaler}, {Morley}, {Morris}, {Mulone}, {Muraveva}, {Musella},
  {Narbonne}, {Nelemans}, {Nicastro}, {Noval}, {Ord{\'e}novic},
  {Ordieres-Mer{\'e}}, {Osborne}, {Pagani}, {Pagano}, {Pailler}, {Palacin},
  {Palaversa}, {Parsons}, {Paulsen}, {Pecoraro}, {Pedrosa}, {Pentik{\"a}inen},
  {Pereira}, {Pichon}, {Piersimoni}, {Pineau}, {Plachy}, {Plum}, {Poujoulet},
  {Pr{\v{s}}a}, {Pulone}, {Ragaini}, {Rago}, {Rambaux}, {Ramos-Lerate},
  {Ranalli}, {Rauw}, {Read}, {Regibo}, {Renk}, {Reyl{\'e}}, {Ribeiro},
  {Rimoldini}, {Ripepi}, {Riva}, {Rixon}, {Roelens}, {Romero-G{\'o}mez},
  {Rowell}, {Royer}, {Rudolph}, {Ruiz-Dern}, {Sadowski}, {Sagrist{\`a}
  Sell{\'e}s}, {Sahlmann}, {Salgado}, {Salguero}, {Sarasso}, {Savietto},
  {Schnorhk}, {Schultheis}, {Sciacca}, {Segol}, {Segovia}, {Segransan},
  {Serpell}, {Shih}, {Smareglia}, {Smart}, {Smith}, {Solano}, {Solitro},
  {Sordo}, {Soria Nieto}, {Souchay}, {Spagna}, {Spoto}, {Stampa}, {Steele},
  {Steidelm{\"u}ller}, {Stephenson}, {Stoev}, {Suess}, {S{\"u}veges}, {Surdej},
  {Szabados}, {Szegedi-Elek}, {Tapiador}, {Taris}, {Tauran}, {Taylor},
  {Teixeira}, {Terrett}, {Tingley}, {Trager}, {Turon}, {Ulla}, {Utrilla},
  {Valentini}, {van Elteren}, {Van Hemelryck}, {van Leeuwen}, {Varadi},
  {Vecchiato}, {Veljanoski}, {Via}, {Vicente}, {Vogt}, {Voss}, {Votruba},
  {Voutsinas}, {Walmsley}, {Weiler}, {Weingrill}, {Werner}, {Wevers},
  {Whitehead}, {Wyrzykowski}, {Yoldas}, {{\v{Z}}erjal}, {Zucker}, {Zurbach},
  {Zwitter}, {Alecu}, {Allen}, {Allende Prieto}, {Amorim},
  {Anglada-Escud{\'e}}, {Arsenijevic}, {Azaz}, {Balm}, {Beck}, {Bernstein},
  {Bigot}, {Bijaoui}, {Blasco}, {Bonfigli}, {Bono}, {Boudreault}, {Bressan},
  {Brown}, {Brunet}, {Bunclark}, {Buonanno}, {Butkevich}, {Carret}, {Carrion},
  {Chemin}, {Ch{\'e}reau}, {Corcione}, {Darmigny}, {de Boer}, {de Teodoro}, {de
  Zeeuw}, {Delle Luche}, {Domingues}, {Dubath}, {Fodor}, {Fr{\'e}zouls},
  {Fries}, {Fustes}, {Fyfe}, {Gallardo}, {Gallegos}, {Gardiol}, {Gebran},
  {Gomboc}, {G{\'o}mez}, {Grux}, {Gueguen}, {Heyrovsky}, {Hoar}, {Iannicola},
  {Isasi Parache}, {Janotto}, {Joliet}, {Jonckheere}, {Keil}, {Kim},
  {Klagyivik}, {Klar}, {Knude}, {Kochukhov}, {Kolka}, {Kos}, {Kutka}, {Lainey},
  {LeBouquin}, {Liu}, {Loreggia}, {Makarov}, {Marseille}, {Martayan},
  {Martinez-Rubi}, {Massart}, {Meynadier}, {Mignot}, {Munari}, {Nguyen},
  {Nordlander}, {Ocvirk}, {O'Flaherty}, {Olias Sanz}, {Ortiz}, {Osorio},
  {Oszkiewicz}, {Ouzounis}, {Palmer}, {Park}, {Pasquato}, {Peltzer}, {Peralta},
  {P{\'e}turaud}, {Pieniluoma}, {Pigozzi}, {Poels}, {Prat}, {Prod'homme},
  {Raison}, {Rebordao}, {Risquez}, {Rocca-Volmerange}, {Rosen}, {Ruiz-Fuertes},
  {Russo}, {Sembay}, {Serraller Vizcaino}, {Short}, {Siebert}, {Silva},
  {Sinachopoulos}, {Slezak}, {Soffel}, {Sosnowska}, {Strai{\v{z}}ys}, {ter
  Linden}, {Terrell}, {Theil}, {Tiede}, {Troisi}, {Tsalmantza}, {Tur},
  {Vaccari}, {Vachier}, {Valles}, {Van Hamme}, {Veltz}, {Virtanen}, {Wallut},
  {Wichmann}, {Wilkinson}, {Ziaeepour}, \& {Zschocke}}]{2016A&A...595A...1G}
{Gaia Collaboration}, {Prusti}, T., {de Bruijne}, J.~H.~J., {et~al.} 2016,
  \aap, 595, A1

\bibitem[{{Gaia Collaboration} {et~al.}(2023){Gaia Collaboration}, {Vallenari},
  {Brown}, {Prusti}, {de Bruijne}, {Arenou}, {Babusiaux}, {Biermann},
  {Creevey}, {Ducourant}, {Evans}, {Eyer}, {Guerra}, {Hutton}, {Jordi},
  {Klioner}, {Lammers}, {Lindegren}, {Luri}, {Mignard}, {Panem}, {Pourbaix},
  {Randich}, {Sartoretti}, {Soubiran}, {Tanga}, {Walton}, {Bailer-Jones},
  {Bastian}, {Drimmel}, {Jansen}, {Katz}, {Lattanzi}, {van Leeuwen}, {Bakker},
  {Cacciari}, {Casta{\~n}eda}, {De Angeli}, {Fabricius}, {Fouesneau},
  {Fr{\'e}mat}, {Galluccio}, {Guerrier}, {Heiter}, {Masana}, {Messineo},
  {Mowlavi}, {Nicolas}, {Nienartowicz}, {Pailler}, {Panuzzo}, {Riclet}, {Roux},
  {Seabroke}, {Sordo}, {Th{\'e}venin}, {Gracia-Abril}, {Portell}, {Teyssier},
  {Altmann}, {Andrae}, {Audard}, {Bellas-Velidis}, {Benson}, {Berthier},
  {Blomme}, {Burgess}, {Busonero}, {Busso}, {C{\'a}novas}, {Carry}, {Cellino},
  {Cheek}, {Clementini}, {Damerdji}, {Davidson}, {de Teodoro}, {Nu{\~n}ez
  Campos}, {Delchambre}, {Dell'Oro}, {Esquej}, {Fern{\'a}ndez-Hern{\'a}ndez},
  {Fraile}, {Garabato}, {Garc{\'\i}a-Lario}, {Gosset}, {Haigron}, {Halbwachs},
  {Hambly}, {Harrison}, {Hern{\'a}ndez}, {Hestroffer}, {Hodgkin}, {Holl},
  {Jan{\ss}en}, {Jevardat de Fombelle}, {Jordan}, {Krone-Martins}, {Lanzafame},
  {L{\"o}ffler}, {Marchal}, {Marrese}, {Moitinho}, {Muinonen}, {Osborne},
  {Pancino}, {Pauwels}, {Recio-Blanco}, {Reyl{\'e}}, {Riello}, {Rimoldini},
  {Roegiers}, {Rybizki}, {Sarro}, {Siopis}, {Smith}, {Sozzetti}, {Utrilla},
  {van Leeuwen}, {Abbas}, {{\'A}brah{\'a}m}, {Abreu Aramburu}, {Aerts},
  {Aguado}, {Ajaj}, {Aldea-Montero}, {Altavilla}, {{\'A}lvarez}, {Alves},
  {Anders}, {Anderson}, {Anglada Varela}, {Antoja}, {Baines}, {Baker},
  {Balaguer-N{\'u}{\~n}ez}, {Balbinot}, {Balog}, {Barache}, {Barbato},
  {Barros}, {Barstow}, {Bartolom{\'e}}, {Bassilana}, {Bauchet}, {Becciani},
  {Bellazzini}, {Berihuete}, {Bernet}, {Bertone}, {Bianchi}, {Binnenfeld},
  {Blanco-Cuaresma}, {Blazere}, {Boch}, {Bombrun}, {Bossini}, {Bouquillon},
  {Bragaglia}, {Bramante}, {Breedt}, {Bressan}, {Brouillet}, {Brugaletta},
  {Bucciarelli}, {Burlacu}, {Butkevich}, {Buzzi}, {Caffau}, {Cancelliere},
  {Cantat-Gaudin}, {Carballo}, {Carlucci}, {Carnerero}, {Carrasco},
  {Casamiquela}, {Castellani}, {Castro-Ginard}, {Chaoul}, {Charlot}, {Chemin},
  {Chiaramida}, {Chiavassa}, {Chornay}, {Comoretto}, {Contursi}, {Cooper},
  {Cornez}, {Cowell}, {Crifo}, {Cropper}, {Crosta}, {Crowley}, {Dafonte},
  {Dapergolas}, {David}, {David}, {de Laverny}, {De Luise}, {De March}, {De
  Ridder}, {de Souza}, {de Torres}, {del Peloso}, {del Pozo}, {Delbo},
  {Delgado}, {Delisle}, {Demouchy}, {Dharmawardena}, {Di Matteo}, {Diakite},
  {Diener}, {Distefano}, {Dolding}, {Edvardsson}, {Enke}, {Fabre}, {Fabrizio},
  {Faigler}, {Fedorets}, {Fernique}, {Fienga}, {Figueras}, {Fournier},
  {Fouron}, {Fragkoudi}, {Gai}, {Garcia-Gutierrez}, {Garcia-Reinaldos},
  {Garc{\'\i}a-Torres}, {Garofalo}, {Gavel}, {Gavras}, {Gerlach}, {Geyer},
  {Giacobbe}, {Gilmore}, {Girona}, {Giuffrida}, {Gomel}, {Gomez},
  {Gonz{\'a}lez-N{\'u}{\~n}ez}, {Gonz{\'a}lez-Santamar{\'\i}a},
  {Gonz{\'a}lez-Vidal}, {Granvik}, {Guillout}, {Guiraud},
  {Guti{\'e}rrez-S{\'a}nchez}, {Guy}, {Hatzidimitriou}, {Hauser}, {Haywood},
  {Helmer}, {Helmi}, {Sarmiento}, {Hidalgo}, {Hilger}, {H{\l}adczuk}, {Hobbs},
  {Holland}, {Huckle}, {Jardine}, {Jasniewicz}, {Jean-Antoine Piccolo},
  {Jim{\'e}nez-Arranz}, {Jorissen}, {Juaristi Campillo}, {Julbe}, {Karbevska},
  {Kervella}, {Khanna}, {Kontizas}, {Kordopatis}, {Korn}, {K{\'o}sp{\'a}l},
  {Kostrzewa-Rutkowska}, {Kruszy{\'n}ska}, {Kun}, {Laizeau}, {Lambert},
  {Lanza}, {Lasne}, {Le Campion}, {Lebreton}, {Lebzelter}, {Leccia}, {Leclerc},
  {Lecoeur-Taibi}, {Liao}, {Licata}, {Lindstr{\o}m}, {Lister}, {Livanou},
  {Lobel}, {Lorca}, {Loup}, {Madrero Pardo}, {Magdaleno Romeo}, {Managau},
  {Mann}, {Manteiga}, {Marchant}, {Marconi}, {Marcos}, {Marcos Santos},
  {Mar{\'\i}n Pina}, {Marinoni}, {Marocco}, {Marshall}, {Martin Polo},
  {Mart{\'\i}n-Fleitas}, {Marton}, {Mary}, {Masip}, {Massari},
  {Mastrobuono-Battisti}, {Mazeh}, {McMillan}, {Messina}, {Michalik}, {Millar},
  {Mints}, {Molina}, {Molinaro}, {Moln{\'a}r}, {Monari}, {Mongui{\'o}},
  {Montegriffo}, {Montero}, {Mor}, {Mora}, {Morbidelli}, {Morel}, {Morris},
  {Muraveva}, {Murphy}, {Musella}, {Nagy}, {Noval}, {Oca{\~n}a}, {Ogden},
  {Ordenovic}, {Osinde}, {Pagani}, {Pagano}, {Palaversa}, {Palicio},
  {Pallas-Quintela}, {Panahi}, {Payne-Wardenaar}, {Pe{\~n}alosa Esteller},
  {Penttil{\"a}}, {Pichon}, {Piersimoni}, {Pineau}, {Plachy}, {Plum}, {Poggio},
  {Pr{\v{s}}a}, {Pulone}, {Racero}, {Ragaini}, {Rainer}, {Raiteri}, {Rambaux},
  {Ramos}, {Ramos-Lerate}, {Re Fiorentin}, {Regibo}, {Richards}, {Rios Diaz},
  {Ripepi}, {Riva}, {Rix}, {Rixon}, {Robichon}, {Robin}, {Robin}, {Roelens},
  {Rogues}, {Rohrbasser}, {Romero-G{\'o}mez}, {Rowell}, {Royer}, {Ruz Mieres},
  {Rybicki}, {Sadowski}, {S{\'a}ez N{\'u}{\~n}ez}, {Sagrist{\`a} Sell{\'e}s},
  {Sahlmann}, {Salguero}, {Samaras}, {Sanchez Gimenez}, {Sanna},
  {Santove{\~n}a}, {Sarasso}, {Schultheis}, {Sciacca}, {Segol}, {Segovia},
  {S{\'e}gransan}, {Semeux}, {Shahaf}, {Siddiqui}, {Siebert}, {Siltala},
  {Silvelo}, {Slezak}, {Slezak}, {Smart}, {Snaith}, {Solano}, {Solitro},
  {Souami}, {Souchay}, {Spagna}, {Spina}, {Spoto}, {Steele},
  {Steidelm{\"u}ller}, {Stephenson}, {S{\"u}veges}, {Surdej}, {Szabados},
  {Szegedi-Elek}, {Taris}, {Taylor}, {Teixeira}, {Tolomei}, {Tonello}, {Torra},
  {Torra}, {Torralba Elipe}, {Trabucchi}, {Tsounis}, {Turon}, {Ulla}, {Unger},
  {Vaillant}, {van Dillen}, {van Reeven}, {Vanel}, {Vecchiato}, {Viala},
  {Vicente}, {Voutsinas}, {Weiler}, {Wevers}, {Wyrzykowski}, {Yoldas}, {Yvard},
  {Zhao}, {Zorec}, {Zucker}, \& {Zwitter}}]{2023A&A...674A...1G}
{Gaia Collaboration}, {Vallenari}, A., {Brown}, A.~G.~A., {et~al.} 2023, \aap,
  674, A1

\bibitem[{{Guinan} \& {Bradstreet}(1988)}]{1988ASIC..241..345G}
{Guinan}, E.~F., \& {Bradstreet}, D.~H. 1988, in NATO Advanced Study Institute
  (ASI) Series C, Vol. 241, Formation and Evolution of Low Mass Stars, ed.
  A.~K. {Dupree} \& M.~T.~V.~T. {Lago}, 345

\bibitem[{{He} {et~al.}(2015){He}, {Zhang}, {Ren}, \&
  {Sun}}]{2015arXiv150201852H}
{He}, K., {Zhang}, X., {Ren}, S., \& {Sun}, J. 2015, arXiv e-prints,
  arXiv:1502.01852, \dodoi{10.48550/arXiv.1502.01852}

\bibitem[{{Jayasinghe} {et~al.}(2018){Jayasinghe}, {Kochanek}, {Stanek},
  {Shappee}, {Holoien}, {Thompson}, {Prieto}, {Dong}, {Pawlak}, {Shields},
  {Pojmanski}, {Otero}, {Britt}, \& {Will}}]{2018MNRAS.477.3145J}
{Jayasinghe}, T., {Kochanek}, C.~S., {Stanek}, K.~Z., {et~al.} 2018, \mnras,
  477, 3145, \dodoi{10.1093/mnras/sty838}

\bibitem[{{Jayasinghe} {et~al.}(2019){Jayasinghe}, {Stanek}, {Kochanek},
  {Shappee}, {Holoien}, {Thompson}, {Prieto}, {Dong}, {Pawlak}, {Pejcha},
  {Shields}, {Pojmanski}, {Otero}, {Hurst}, {Britt}, \&
  {Will}}]{2019MNRAS.485..961J}
{Jayasinghe}, T., {Stanek}, K.~Z., {Kochanek}, C.~S., {et~al.} 2019, \mnras,
  485, 961, \dodoi{10.1093/mnras/stz444}

\bibitem[{{Jayasinghe} {et~al.}(2020){Jayasinghe}, {Stanek}, {Kochanek},
  {Shappee}, {Pinsonneault}, {Holoien}, {Thompson}, {Prieto}, {Pawlak},
  {Pejcha}, {Pojmanski}, {Otero}, {Hurst}, \& {Will}}]{2020MNRAS.493.4045J}
---. 2020, \mnras, 493, 4045, \dodoi{10.1093/mnras/staa518}

\bibitem[{{Kingma} \& {Ba}(2014)}]{2014arXiv1412.6980K}
{Kingma}, D.~P., \& {Ba}, J. 2014, arXiv e-prints, arXiv:1412.6980,
  \dodoi{10.48550/arXiv.1412.6980}

\bibitem[{{Kuiper}(1941)}]{1941ApJ....93..133K}
{Kuiper}, G.~P. 1941, \apj, 93, 133, \dodoi{10.1086/144252}

\bibitem[{{Kwee} \& {van Woerden}(1956)}]{1956BAN....12..327K}
{Kwee}, K.~K., \& {van Woerden}, H. 1956, \bain, 12, 327

\bibitem[{{Lafler} \& {Kinman}(1965)}]{1965ApJS...11..216L}
{Lafler}, J., \& {Kinman}, T.~D. 1965, \apjs, 11, 216, \dodoi{10.1086/190116}

\bibitem[{{Li} {et~al.}(2022){Li}, {Gao}, {Liu}, {Gao}, {Li}, {Chen}, \&
  {Sun}}]{2022AJ....164..202L}
{Li}, K., {Gao}, X., {Liu}, X.-Y., {et~al.} 2022, \aj, 164, 202

\bibitem[{{Li} {et~al.}(2020){Li}, {Kim}, {Xia}, {Michel}, {Hu}, {Gao}, {Guo},
  \& {Chen}}]{2020AJ....159..189L}
{Li}, K., {Kim}, C.-H., {Xia}, Q.-Q., {et~al.} 2020, \aj, 159, 189

\bibitem[{{Li} {et~al.}(2019){Li}, {Xia}, {Michel}, {Hu}, {Guo}, {Gao}, {Chen},
  \& {Gao}}]{2019MNRAS.485.4588L}
{Li}, K., {Xia}, Q.-Q., {Michel}, R., {et~al.} 2019, \mnras, 485, 4588

\bibitem[{{Li} {et~al.}(2024{\natexlab{a}}){Li}, {Gao}, {Guo}, {Gao}, {Chen},
  {Wang}, {Xin}, {Han}, {Kim}, \& {Jeong}}]{2024A&A...692L...4L}
{Li}, K., {Gao}, X., {Guo}, D.-F., {et~al.} 2024{\natexlab{a}}, \aap, 692, L4,
  \dodoi{10.1051/0004-6361/202451947}

\bibitem[{{Li} {et~al.}(2024{\natexlab{b}}){Li}, {Zhu}, {Ding}, {Xu}, {Zheng},
  {Qiu}, \& {Liu}}]{2024ApJS..271...32L}
{Li}, X.-Z., {Zhu}, Q.-F., {Ding}, X., {et~al.} 2024{\natexlab{b}}, \apjs, 271,
  32, \dodoi{10.3847/1538-4365/ad226a}

\bibitem[{{Liu}(2021)}]{2021PASP..133h4202L}
{Liu}, L. 2021, \pasp, 133, 084202, \dodoi{10.1088/1538-3873/ac1ac1}

\bibitem[{{Lucy}(1967)}]{1967ZA.....65...89L}
{Lucy}, L.~B. 1967, \zap, 65, 89

\bibitem[{{Lucy}(1968{\natexlab{a}})}]{1968ApJ...151.1123L}
---. 1968{\natexlab{a}}, \apj, 151, 1123, \dodoi{10.1086/149510}

\bibitem[{{Lucy}(1968{\natexlab{b}})}]{1968ApJ...153..877L}
---. 1968{\natexlab{b}}, \apj, 153, 877, \dodoi{10.1086/149712}

\bibitem[{{Mowlavi} {et~al.}(2023){Mowlavi}, {Holl}, {Lecoeur-Ta{\"\i}bi},
  {Barblan}, {Kochoska}, {Pr{\v{s}}a}, {Mazeh}, {Rimoldini}, {Gavras},
  {Audard}, {Jevardat de Fombelle}, {Nienartowicz}, {Garc{\'\i}a-Lario}, \&
  {Eyer}}]{2023A&A...674A..16M}
{Mowlavi}, N., {Holl}, B., {Lecoeur-Ta{\"\i}bi}, I., {et~al.} 2023, \aap, 674,
  A16, \dodoi{10.1051/0004-6361/202245330}

\bibitem[{{O'Connell}(1951)}]{1951PRCO....2...85O}
{O'Connell}, D.~J.~K. 1951, Publications of the Riverview College Observatory,
  2, 85

\bibitem[{{Pr{\v{s}}a} \& {Zwitter}(2005)}]{2005ApJ...628..426P}
{Pr{\v{s}}a}, A., \& {Zwitter}, T. 2005, \apj, 628, 426, \dodoi{10.1086/430591}

\bibitem[{{Pr{\v{s}}a} {et~al.}(2016){Pr{\v{s}}a}, {Conroy}, {Horvat}, {Pablo},
  {Kochoska}, {Bloemen}, {Giammarco}, {Hambleton}, \&
  {Degroote}}]{2016ApJS..227...29P}
{Pr{\v{s}}a}, A., {Conroy}, K.~E., {Horvat}, M., {et~al.} 2016, \apjs, 227, 29,
  \dodoi{10.3847/1538-4365/227/2/29}

\bibitem[{{Pr{\v{s}}a} {et~al.}(2022){Pr{\v{s}}a}, {Kochoska}, {Conroy},
  {Eisner}, {Hey}, {IJspeert}, {Kruse}, {Fleming}, {Johnston}, {Kristiansen},
  {LaCourse}, {Mortensen}, {Pepper}, {Stassun}, {Torres}, {Abdul-Masih},
  {Chakraborty}, {Gagliano}, {Guo}, {Hambleton}, {Hong}, {Jacobs}, {Jones},
  {Kostov}, {Lee}, {Omohundro}, {Orosz}, {Page}, {Powell}, {Rappaport}, {Reed},
  {Schnittman}, {Schwengeler}, {Shporer}, {Terentev}, {Vanderburg}, {Welsh},
  {Caldwell}, {Doty}, {Jenkins}, {Latham}, {Ricker}, {Seager}, {Schlieder},
  {Shiao}, {Vanderspek}, \& {Winn}}]{2022ApJS..258...16P}
{Pr{\v{s}}a}, A., {Kochoska}, A., {Conroy}, K.~E., {et~al.} 2022, \apjs, 258,
  16, \dodoi{10.3847/1538-4365/ac324a}

\bibitem[{{Qian} {et~al.}(2017){Qian}, {He}, {Zhang}, {Zhu}, {Shi}, {Zhao}, \&
  {Zhou}}]{2017RAA....17...87Q}
{Qian}, S.-B., {He}, J.-J., {Zhang}, J., {et~al.} 2017, Research in Astronomy
  and Astrophysics, 17, 087

\bibitem[{{Qian} {et~al.}(2007){Qian}, {Liu}, {Soonthornthum}, {Zhu}, \&
  {He}}]{2007AJ....134.1475Q}
{Qian}, S.~B., {Liu}, L., {Soonthornthum}, B., {Zhu}, L.~Y., \& {He}, J.~J.
  2007, \aj, 134, 1475, \dodoi{10.1086/521432}

\bibitem[{{Rasio}(1995)}]{1995ApJ...444L..41R}
{Rasio}, F.~A. 1995, \apjl, 444, L41

\bibitem[{{Ruci{\'n}ski}(1969)}]{1969AcA....19..245R}
{Ruci{\'n}ski}, S.~M. 1969, \actaa, 19, 245

\bibitem[{{Rucinski}(1992)}]{1992AJ....103..960R}
{Rucinski}, S.~M. 1992, \aj, 103, 960

\bibitem[{{Rucinski}(1994)}]{1994PASP..106..462R}
---. 1994, \pasp, 106, 462

\bibitem[{{Rucinski}(2007)}]{2007MNRAS.382..393R}
---. 2007, \mnras, 382, 393

\bibitem[{{Rucinski} {et~al.}(2001){Rucinski}, {Lu}, {Mochnacki}, {Og{\l}oza},
  \& {Stachowski}}]{2001AJ....122.1974R}
{Rucinski}, S.~M., {Lu}, W., {Mochnacki}, S.~W., {Og{\l}oza}, W., \&
  {Stachowski}, G. 2001, \aj, 122, 1974, \dodoi{10.1086/323106}

\bibitem[{{Shappee} {et~al.}(2014){Shappee}, {Prieto}, {Grupe}, {Kochanek},
  {Stanek}, {De Rosa}, {Mathur}, {Zu}, {Peterson}, {Pogge}, {Komossa}, {Im},
  {Jencson}, {Holoien}, {Basu}, {Beacom}, {Szczygie{\l}}, {Brimacombe},
  {Adams}, {Campillay}, {Choi}, {Contreras}, {Dietrich}, {Dubberley},
  {Elphick}, {Foale}, {Giustini}, {Gonzalez}, {Hawkins}, {Howell}, {Hsiao},
  {Koss}, {Leighly}, {Morrell}, {Mudd}, {Mullins}, {Nugent}, {Parrent},
  {Phillips}, {Pojmanski}, {Rosing}, {Ross}, {Sand}, {Terndrup}, {Valenti},
  {Walker}, \& {Yoon}}]{2014ApJ...788...48S}
{Shappee}, B.~J., {Prieto}, J.~L., {Grupe}, D., {et~al.} 2014, \apj, 788, 48

\bibitem[{{Sun} {et~al.}(2020){Sun}, {Chen}, {Deng}, \& {de
  Grijs}}]{2020ApJS..247...50S}
{Sun}, W., {Chen}, X., {Deng}, L., \& {de Grijs}, R. 2020, \apjs, 247, 50,
  \dodoi{10.3847/1538-4365/ab7894}

\bibitem[{{Wang} {et~al.}(2024){Wang}, {Ding}, {Li}, {Xiong}, {Cheng}, \&
  {Ji}}]{2024ApJS..273...31W}
{Wang}, J., {Ding}, X., {Li}, J., {et~al.} 2024, \apjs, 273, 31,
  \dodoi{10.3847/1538-4365/ad5953}

\bibitem[{{Wilson}(1979)}]{1979ApJ...234.1054W}
{Wilson}, R.~E. 1979, \apj, 234, 1054

\bibitem[{{Wilson}(1990)}]{1990ApJ...356..613W}
---. 1990, \apj, 356, 613

\bibitem[{{Wilson} \& {Devinney}(1971)}]{1971ApJ...166..605W}
{Wilson}, R.~E., \& {Devinney}, E.~J. 1971, \apj, 166, 605

\bibitem[{{Zhang}(2024)}]{2024NatSR..1413011Z}
{Zhang}, X.-D. 2024, Scientific Reports, 14, 13011

\end{thebibliography}
\bibliographystyle{aasjournal}



\end{document}